\documentclass[english]{article}
\usepackage{amsmath}
\usepackage{amssymb}
\usepackage{cite}
\usepackage{hyperref}
\usepackage{graphicx}
\usepackage{caption}
\usepackage{subcaption}
\usepackage{tikz}
\usetikzlibrary{decorations.pathreplacing}
\usetikzlibrary{shapes}
\makeatletter

\usepackage{babel}

\begin{document}
\begin{titlepage}
\thispagestyle{empty}

\bigskip

\begin{center}
\noindent{\Large \textbf
{Non-Integrability of Strings in $AdS_{6}\times S^{2}\times\Sigma$ Background and its 5D Holographic Duals\\}}

\vspace{0,5cm}

\noindent{G. Alencar ${}^{a,c}$\footnote{e-mail: geova@fisica.ufc.br}, M. O. Tahim ${}^{b,c}$}

\vspace{0,5cm}
 
  {\it ${}^a$Departamento de F\'{\i}sica, Universidade Federal do Cear\'{a}-
Caixa Postal 6030, Campus do Pici, 60455-760, Fortaleza, Cear\'{a}, Brazil. 
 }

\vspace{0.2cm}
{\it ${}^b$Universidade Estadual do Cear\'a, Faculdade de Educa\c c\~ao, Ci\^encias e Letras do Sert\~ao Central- 
R. Epitácio Pessoa, 2554, 63.900-000  Quixad\'{a}, Cear\'{a},  Brazil.
 
 \vspace{0.2cm}
{\it ${}^c$International Institute of Physics - Federal University of Rio Grande do Norte, Campus Universit\'ario, Lagoa Nova, Natal, RN 59078-970, Brazil.
 }

 }

\end{center}

\vspace{0.3cm}
\begin{abstract}
In this manuscript we study Liouvillian non-integrability of strings in $AdS_{6}\times S^{2}\times\Sigma$ background. We consider soliton strings and look for simple solutions in order to reduce the equations to only one linear second order differential equation called Normal Variation Equation (NVE). We study truncations in $\eta$ and $\sigma$ variables showing their applicability or not to catch (non) integrability of models. With this technique we are able to study many recent cases considered in the literature: the abelian and non-abelian T-duals, the $(p,q)$-5-brane system, the T$_{N}$, $+_{MN}$  theories and the $\tilde{T}_{N,P}$ and $+_{P,N}$ quivers. We show that all of them are not integrable. Finally, we consider the general case at the boundary $\sigma=\sigma_0$ for large $\sigma_0$ and show that we can get general conclusions about integrability. For example, beyond the above quivers, we show generically that long quivers are not integrable. In order to establish the results, we numerically study the string dynamical system seeking by chaotic behaviour. Such a characteristic gives one more piece of evidence for non-integrability of the background studied.
\end{abstract}

\end{titlepage}

\tableofcontents

\newpage
\section{Introduction.}

In a recent paper, in the context of the $AdS/CFT$ correspondence \cite{Maldacena:1997re}, a precise relation between type II string theory and $d=5$ conformal field theories was described \cite{Legramandi:2021uds}. Given the $AdS$ space-time string background, when the BPS equations ensuring half-SUSY and Bianchi identity are imposed, a potential function obeying a linear PDE (a Laplace equation) is found with proper boundary conditions. This potential contains all information about the background and, because of the quantization of Page charges, one of the boundary conditions implies that it is written in terms of a \textit{Rank function}, a convex polygonal function, linear by pieces, with integer values at integer points. Then, this Rank Function is put in contact with a quiver gauge theory: the ranks of the gauge and flavour groups are organized in it. Because the Laplace equation is central in this description, it is referred as an \textit{electrostatic description}. In the research for half-BPS backgrounds of the type $AdS_{D}\times S^{2}\times \Sigma_{8-D}$ the same steps can be built for several dimensions $D$ \cite{Akhond:2021ffz} (see references therein). After the field theory string background pair is identified, various tests and predictions of the correspondence follows in a simpler way. In this work we want to take profit of this formalism in order to study Liouvillian (non) integrability of strings in the $AdS_{6}\times S^{2}\times\Sigma$ background.

The $AdS_{6}\times S^{2}\times\Sigma$ background was in fact found in \cite{DHoker:2016ujz} and, after this, the correspondence was tested in a long list of papers
\cite{DHoker:2016ysh,DHoker:2017mds,Gutperle:2017tjo,DHoker:2017zwj,Gutperle:2018vdd,Fluder:2018chf,Bergman:2018hin,Lozano:2018pcp,Uhlemann:2019ypp,Aharony:1997bh,Benini:2009gi,Bergman:2014kza,Hayashi:2014hfa,Uhlemann:2020bek}. The fact that the dual CFT is not unique has allowed the construction of a lot of 5D quiver theories: $+_{N,M}$ in Ref. \cite{Aharony:1997bh}, $T_N$ in Refs. \cite{Benini:2009gi,Bergman:2014kza,Hayashi:2014hfa}, $Y_N$, $\ensuremath\diagup\!\!\!\!\!{+}_N$ and $+_{N,M,j}$ in Ref. \cite{Bergman:2018hin}, $T_{2K,K,2}$ and $T_{N,K,j}$ in Ref. \cite{Chaney:2018gjc} and many others (see \cite{Uhlemann:2019ypp} and references therein). Beyond this we can also cite the abelian and non-abelian T-dual backgrounds, in which the dual CFT is not well known \cite{Lozano:2018pcp, Lozano:2012au,Lozano:2013oma}. 

As the correspondence signals, several characteristics should match in the string side and in gauge theory side. An important characteristic is just the integrability of the model: it helps us in using powerful techniques to study conjectured relations nonperturbatively. In the case of the $AdS_{5}\times S^{5}$ it is already known the existence of integrability structures behind it.  The study of integrability, from a classical perspective, can be made by following some specific strategies. One of them is by searching for existence/non-existence of chaos in the associated dynamical system given by the string model: by finding chaotic behaviour (studying, for example, the Lyapunov exponents and Poincare section`s structure), integrability is excluded. There is an good number of references of chaos research in string related topics \cite{Zayas:2010fs,Basu:2011dg,Basu:2011fw,Basu:2012ae,Stepanchuk:2012xi,Giataganas:2014hma,Chervonyi:2013eja,Asano:2015eha,
Asano:2015qwa,Ishii:2016rlk,Panigrahi:2016zny,Ma:2019ewq,Basu:2016zkr,Roychowdhury:2017vdo,Nunez:2018ags,Nunez:2018qcj,Rigatos:2020hlq,Rigatos:2020igd
}. Another path is to look for a Lax pair formulation of the string model. Lax pairs, if they exist for a classical system, can be used to generate a tower of integrals of motion and this will give support to integrability. In the case of string theory, this approach is very successful when the string background is coset-like, as in the case of $AdS_{5}\times S^{5}$ \cite{Bena:2003wd,Alday:2005gi,Hoare:2015gda} and the simple case of $R\times S^{3}$ \cite{Vicedo:2011zz}. But in fact, there is no general guide to find a Lax pair formulation. 

The last path we would like to cite is an analytical method to discuss the existence of integrability in a dynamical system. The idea is to find a string soliton and show that the dynamics of such an object is (non) integrable in some sense. This method has been recently used to study integrability of a lot of string backgrounds, their respective duals and in other models \cite{Roychowdhury:2017vdo,Nunez:2018ags,Nunez:2018qcj,Filippas:2019puw,Roychowdhury:2020abu,Roychowdhury:2020zer,Pal:2021fzb}. Given a system of differential equations, the analysis of the variational equation around a particular solution can show its (non) integrability. In other words, if a nonlinear system admits first integrals, the variational equation will admit it too. Disproving this for a given class of functions will imply in the non integrability of the initial nonlinear system. The mathematical establishment of integrability through the normal variation equation (NVE) has been made by some tests that were improved along the years. First, there is  Ziglin`s theorem relating the existence of first integral of motion with monodromy matrices around the straight line solution, the basis to linearize the system of differential equations \cite{Ziglin:1983fs,Ziglin:1983sc}. After that, techniques of differential Galois` theory applied to the NVE equation were introduced \cite{Simo:1994ab,Ramis:2007ab,Ramis:2001ab}. In this work we make use of the improvement made by Kovacic \cite{Kovacic85}.  It gives, through an specific algorithm, an answer to the existence of integrability: once the NVE is written in a linear form with polynomial coefficients, it suffices to check a group of criteria. In fact, Kovacic provided a way to construct the solutions. In the case of string models, it basically consists in the following: first we  find the equations of motion for the $l$ degrees of freedom of a proposed string soliton. Next, we find simple solutions for  $(l-1)$ of these equations which are replaced in the last one. They give us the normal variation equation (NVE). It is a linear second order differential equation given by 
$$
z''+{\cal B}z'+{\cal A}z=0.
$$
With the equation above at hand, we can use the Kovacic's criteria to seek if a Liouvillian solution do exist. As will be explained, the functions ${\cal A},{\cal B}$ and its derivatives determine the existence of a closed form of Liouvillian solutions. We should point out that the non-integrability in some of these backgrounds has been studied in Ref.\cite{Wulff:2019tzh}. The authors use the assumption that they allow for a regular critical point for the warp factor, which is necessary to have a GKP string solution. With this assumption they were able to show non-integrability for the background of Ref. \cite{Apruzzi:2014qva}. However that background does not include D7 branes and further generalizations \cite{DHoker:2017zwj,Gutperle:2018vdd,Lozano:2018pcp,Chaney:2018gjc}. Namely, the only explicit examples without D7 branes considered here are  the $T_N$ and $+_{MN}$. For these, our results enforce the findings of Ref. \cite{Wulff:2019tzh} about non-integrability. In this paper our results include D7 branes and therefore are more general: we are able to find non integrability for a larger class of backgrounds.

In this work we study analytical and numerical (non) integrability of strings in $AdS_{6}\times S^{2}\times\Sigma$ background in direct connection to its 5D Holographic Duals, as suggested by the electrostatic method which is reviewed in section $2$ . In section $3$ we study the string dynamics in the given background and write the NVE that will be basis for the conclusions. In sections $4$ and $5$ we apply Kovacic's criteria to several potentials, running for regions where $\sigma = \sigma_0$ and $\eta=\eta_0$, including those supporting quiver gauge models. In section $6$ we discuss the potential for a general case, studying how to see and avoid logarithmic terms at the final $U$- function in order to apply Kovacic's criteria. In section $7$ we supplement the results coming from the analytical process by finding chaotic behaviour in the dynamical system. We compute string trajectories, power spectra, Lyapunov exponents and Poincar\'{e} sections in order to establish the results. Finally, we present our conclusions.

\section{The Electrostatic Description}
In this section we quickly review the correspondence between strings in $AdS_{6}\times S^{2}\times\Sigma$ background and $d=5$ SCFT through the electrostatic viewpoint.

\subsection{The Background.}
The type IIB background that is important in this work is described in Ref.\cite{Legramandi:2021uds}. The full configuration consists of the metric, the dilaton, $B_2$, $C_2$ and $C_0$-fields in the NS and Ramond sectors respectively. In string frame it is given by
\begin{eqnarray}
& & ds_{10,st}^2= f_1(\sigma,\eta)\Big[ds^2(\text{AdS}_6) + f_2(\sigma,\eta)ds^2(S^2) + f_3(\sigma,\eta)(d\sigma^2+d\eta^2) \Big],\;\;e^{-2\Phi}=f_6(\sigma,\eta) , \nonumber\\[2mm]
& & B_2=f_4(\sigma,\eta) \text{Vol}(S^2),\;\;C_2= f_5(\sigma,\eta) \text{Vol}(S^2),\;\;\; C_0= f_7(\sigma,\eta), \label{background}\\[2mm]
& & f_1= \frac{3 \pi}{2}\sqrt{\sigma^2 +\frac{3\sigma \partial_\sigma V}{\partial^2_\eta V}},\;\; f_2= \frac{\partial_\sigma V \partial^2_\eta V}{3\Lambda},\;\;f_3= \frac{\partial^2_\eta V}{3\sigma \partial_\sigma V},\;\;\Lambda=\sigma(\partial_\sigma\partial_\eta V)^2 + (\partial_\sigma V-\sigma \partial^2_\sigma V)  \partial^2_\eta V,\nonumber\\[2mm]
& & f_4=\frac{\pi}{2}\left(\eta -\frac{(\sigma \partial_\sigma V) (\partial_\sigma\partial_\eta V)}{\Lambda} \right),\;\;\;\; f_5=\frac{\pi}{2}\left( V- \frac{\sigma\partial_\sigma V}{\Lambda} (\partial_\eta V (\partial_\sigma \partial_\eta V) -3 (\partial^2_\eta V)(\partial_\sigma V)) \right),\nonumber\\[2mm]
& & f_6=12 \frac{\sigma^2 \partial_\sigma V \partial^2_\eta V}{(3 \partial_\sigma V +\sigma \partial^2_\eta V)^2}\Lambda,\;\;\;\; f_7=2\left( \partial_\eta V + \frac{(3\sigma \partial_\sigma V) (\partial_\sigma\partial_\eta V )}{3\partial_\sigma V +\sigma \partial^2_\eta V}  \right).\nonumber
\end{eqnarray}
In the  equations above the range of $\eta$ and $\sigma$ are the intervals $[0,P]$ and $-\infty<\sigma<\infty$ respectively. We also use the following parametrization
$$
ds_{S^{2}\text{ }}^{2}=d\chi^{2}+\sin^{2}\chi d\xi^{2},\qquad\mbox{Vol}(S^{2})=\sin\chi d\chi\wedge d\xi
$$
and
$$
ds_{AdS_{6}}^{2}=d\rho^{2}-\cosh^{2}\rho dt^{2}+\sinh^{2}\rho d\Omega_{4}^{2},
$$
with 
$$
d\Omega_{4}^{2}=d\theta_{1}^{2}+\cos^{2}\theta_{1}\left(d\theta_{2}^{2}+\cos^{2}\theta_{2}d\phi_{1}^{2}+\sin^{2}\theta_{2}d\phi_{2}^{2}\right).
$$

The background depends only of one potential function $V(\sigma,\eta)$, which  solves a linear partial differential equation given by
\begin{equation}
\partial_\sigma \left(\sigma^2 \partial_\sigma V\right) +\sigma^2 \partial^2_\eta V=0.\label{diffeq}
 \end{equation}
Next, we define  
\begin{equation}
V(\sigma,\eta)=\frac{\hat{V} (\sigma,\eta)}{\sigma},\label{change1}
\end{equation}
to arrive at  a Laplace equation given by
\begin{equation}
\partial^2_\sigma \hat{V} + \partial_\eta^2 \hat{V}=0.\label{eqfinal}
\end{equation}
The boundary conditions are 
\begin{eqnarray}
& & \hat{V}(\sigma\to\pm\infty,\eta)=0,\;\;\;\;\;\hat{V}(\sigma, \eta=0)= \hat{V}(\sigma, \eta=P)=0.\nonumber\\
& & \lim_{\epsilon\to 0}\left(\partial_\sigma \hat{V}(\sigma=+\epsilon,\eta)- \partial_\sigma \hat{V}(\sigma=-\epsilon,\eta)\right)= {\cal R}(\eta).\label{bc}
\end{eqnarray}
Due to the Laplace equation  above and boundary conditions, the authors of Ref. \cite{Legramandi:2021uds} called this  an ``electrostatic description'' of the system.

The solution of the equation above is given by 
\begin{equation}\label{solutionVhat}
\hat{V}(\sigma,\eta)=\sum_{k=1}^{\infty}a_{k}\sin\left(\frac{k\pi}{P}\eta\right){e^{-\frac{k\pi}{P}|\sigma|}},\;\;\;\;a_{k}=\frac{1}{\pi k}\int_{0}^{P}{\cal R}(\eta)\sin\left(\frac{k\pi}{P}\eta\right)~d\eta,
\end{equation}
where the rank function is obtained as
\begin{equation}
{\cal R}(\eta)=\sum_{k=1}^{\infty}c_{k}\sin\left(\frac{k\pi}{P}\eta\right),\;\;\;\;\;2\pi ka_{k}=-Pc_{k}.
\end{equation}
${\cal R}(\eta)$ is interpreted as a charge density at $\sigma=0$, extended along $0\leq\eta\leq P$ ($P\in \mathbb{Z}$). Because of the quantization of Page charges, it is fixed to be a convex piece-wise linear function. It is this function that will make possible the precise connection between the string model and the quiver gauge models in the correspondence.
\subsection{The Holographic Duals.}
Now we review the holographic duals of the background just described. For this we follow the lines of Refs. \cite{Legramandi:2021uds} and  \cite{Uhlemann:2019ypp}. In order to find a general solution, the quantization of the Page charges is necessary. This and the boundary conditions (\ref{bc}) enforce the Rank function ${\cal R}(\eta)$ to be given by
 \[ {\cal R}(\eta) = \begin{cases} 
          N_1 \eta & 0\leq \eta \leq 1 \\
          N_l+ (N_{l+1} - N_l)(\eta-l) & l \leq \eta\leq l+1,\;\;\; l:=1,...., P-2\\
  %
          N_{P-1}(P-\eta) & (P-1)\leq \eta\leq P . 
       \end{cases}
    \]
Depending on the choice of the Rank function $\cal{R}$,  the number of $D7,D5$ and $NS5$ branes can be determined. This also fix the strong coupling CFT to which a linear quiver theory flows. The gauge group  of the linear quiver is given by $\Pi_{i=1}^{P-1} SU(N_i)$, with each gauge connected by bifundamental hypermultiplets. Finally, some of the color groups can have $SU(N_f)$ flavor groups. However, the relation to the holographic dual is trustable only in the limit of very large $P$. The only exception to the general solution above are the abelian and non-abelian T-duals, in which the boundary conditions (\ref{bc}) are not satisfied and the dual CFTs are not well known.  In the rest of this section we give the details of some particular solutions we will consider later.

\subsubsection{Abelian and non-Abelian T-Duals.}
The first $AdS_6$ solutions to Type IIB supergravity were first constructed in \cite{Lozano:2012au,Lozano:2013oma} by acting with Abelian and non-Abelian T-duality on the Brandhuber-Oz solution to massive IIA. Later, it was shown to fit in the construction of D'Hoker, Gutperle and Uhlemann\cite{Lozano:2018pcp} and therefore also in the electrostatic description of \cite{Legramandi:2021uds}.  Despite of being the first IIB constructions, the dual theories are not yet well known (see \cite{Lozano:2018pcp} and references therein). 

The abelian case includes $D7/O7$-branes and provides an example of a Riemann surface with the topology of an annulus \cite{Lozano:2018pcp}. The electrostatic description has a potential given by 
\begin{equation}\label{VATD}
V_{ATD}=\frac{b_{1}}{\sigma}-b_{4}(3\eta^{2}-\sigma^{2}),\,b_{1}=\frac{81}{512},\,b_{4}=\frac{m}{486}.
\end{equation}
The non-Abelian case arises from the upper half-plane\cite{Lozano:2013oma}. In the electrostatic description it has potential given by
 \begin{equation}\label{VNATD}
V_{NATD}=\frac{a_{1}\eta}{\sigma}+4a_{4}(\eta\sigma^{2}-\eta^{3}),\,a_{1}=\frac{1}{128},\,a_{4}=\frac{m}{432}.
\end{equation}
These potentials can be obtained as partial polynomial expansions for $\sigma\geq 0$ from the full electrostatic description \cite{Legramandi:2021uds}. We should point that they do not satisfy the boundary conditions (\ref{bc}).

\subsubsection{The $T_N$ and $+_{MN}$ Theories.}
 The electrostatic description for these models were given in Ref.\cite{Legramandi:2021uds}. The $T_N$ theory was first studied in Refs. \cite{Benini:2009gi,Bergman:2014kza,Hayashi:2014hfa}. It is the strongly-coupled UV fixed point of the linear quiver gauge with junctions of $N$ $D5$, $N$ $NS5$ and $N$ $(1,1)$ $5$-branes. Its electrostatic description has potential given by 
\begin{equation}\label{VTN}
\hat{V}=\frac{9N^{2}}{32\pi^{2}}\sum_{k=1}^{\infty}\frac{(-1)^{k+1}}{k^{2}}\sin\left(\frac{4k\pi}{9N}\eta\right)e^{-\frac{4k\pi}{9N}|\sigma|}.
\end{equation}
The $+_{MN}$ theory was already studied in Ref. \cite{Aharony:1997bh} and is defined on the intersection of $N$ $D5$ and $M$ $NS5$-branes. In field theory it can be defined as the UV fixed point of the linear quiver gauge theory. In this case, the electrostatic description has potential given by
\begin{equation}\label{V+MN}
\hat{V}=\frac{9MN}{32\pi^{2}}\sum_{k=1}^{\infty}\frac{1-(-1)^{k}}{k^{2}}\sin\left(\frac{4\pi k}{9M}\eta\right)e^{-\frac{4\pi k}{9M}|\sigma|}.
\end{equation}

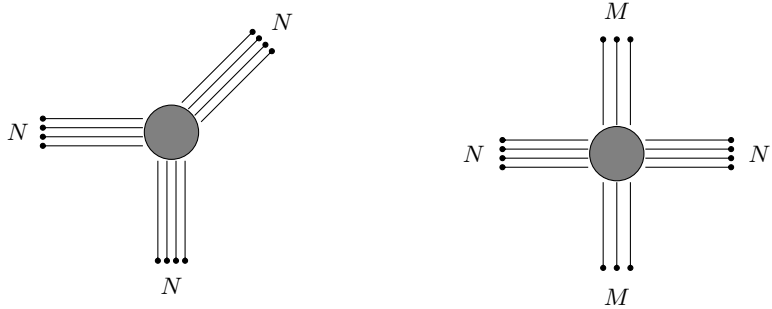
\begin{figure}[h!!]
\begin{center}
\begin{subfigure}{.5\textwidth}
 \begin{tikzpicture}[scale=0.95]
    \foreach \i in {-3/2,-1/2,1/2,3/2}{
      \draw (-0.4,0.127*\i) -- (-1.8,0.127*\i) [fill=black] circle (1pt);
      \draw (0.127*\i,-0.4) -- (0.127*\i,-1.8) [fill=black] circle (1pt);
      \draw (0.28+0.09*\i,0.28-0.09*\i) -- (1.27+0.09*\i,1.27-0.09*\i) [fill=black] circle (1pt) ;
    }
    \draw[fill=gray] (0,0) circle (0.38);    
    \node at (0,-2.15) {\small $N$};
    \node at (-2.15,0) {\small $N$};
    \node at (1.55,1.55) {\small $N$};
\end{tikzpicture}
\label{fig:TN}
\end{subfigure}\hskip 0mm
\begin{subfigure}{.5\textwidth}
 \begin{tikzpicture}[scale=0.95]
    \foreach \i in {-3/2,-1/2,1/2,3/2}{
      \draw (-0.4,0.127*\i) -- (-1.6,0.127*\i) [fill=black] circle (1pt);
      \draw (0.4,0.127*\i) -- (1.6,0.127*\i) [fill=black] circle (1pt);
    }
    \foreach \i in {-3/2,0,3/2}{
      \draw (0.127*\i,-0.4) -- (0.127*\i,-1.6) [fill=black] circle (1pt);
      \draw (0.127*\i,0.4) -- (0.127*\i,1.6) [fill=black] circle (1pt);
    }
    \draw[fill=gray] (0,0) circle (0.38);
    \node at (0,-2.0) {\small $M$};
    \node at (2.0,0) {\small $N$};
    \node at (0,2.0) {\small $M$};
    \node at (-2.0,0) {\small $N$};
\end{tikzpicture}
\label{fig:plus}
\end{subfigure}\hskip 0mm
 \caption{Brane diagrams for the a\mbox{)} $T_N$ and b\mbox{)} $+_{N,M}$  theories. The $(p,q)$-$5$-branes are represented by straight lines and 7-branes represented by filled black dots. The angles are determined by the $(p,q)$ charges and the 5d SCFTs are realized by intersections at a point.\label{fig:unconstrained}}
\end{center}
\end{figure}

In Fig. \ref{fig:unconstrained} we give a brane diagram\footnote{Figure adapted from Ref. \cite{Uhlemann:2019ypp}.} in order to visualize the quivers above.  
We should point that both the theories above have $N_f=2N$. For more examples with this number of flavors see Ref. \cite{Uhlemann:2019ypp}.

\subsubsection{The $\tilde{T}_{N,P}$ and $+_{P,N}$ Theories.}
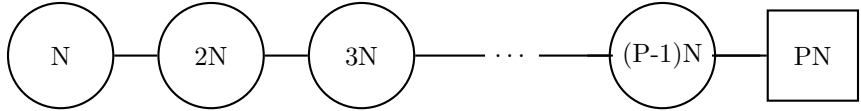
\begin{figure}[h!]
\begin{center}
	\begin{tikzpicture}
	\node (1) at (-6,0) [circle,draw,thick,minimum size=1.4cm] {N};
	\node (2) at (-4,0) [circle,draw,thick,minimum size=1.4cm] {2N};
	\node (3) at (-2,0) [circle,draw,thick,minimum size=1.4cm] {3N};	
	\node (4) at (0,0)  {$\dots$};
	\node (6) at (4,0) [rectangle,draw,thick,minimum size=1.2cm] {PN};
	\node (5) at (2,0)  {(P-1)N};
	\draw[thick] (1) -- (2) -- (3) -- (4) -- (5)-- (6);
	\draw[thick] (2,0) circle (0.7cm) ;
	\draw[thick] (1,0) -- (1.3,0);
	\draw[thick] (2.7,0) -- (3.3,0);
	\end{tikzpicture}
\caption{The $\tilde{T}_{N,P}$ quiver, with $P N$ D7-branes, $NP(P-1)/2$ D5-branes and $P$ NS5-branes.}
\label{FIGTNP}
\end{center}
\end{figure}

These theories were both studied in Ref.\cite{Legramandi:2021uds}. The rank function associated  to $\tilde{T}_{N,P}$ theory  
is given by
\begin{equation}\label{rankTNP}
{\cal R}(\eta) = \begin{cases} 
N\eta & 0\leq \eta \leq (P-1) \\
N(P-1) (P-\eta)& (P-1)\leq \eta\leq P .
\end{cases}
\end{equation}
The number of D7-branes is given by $P N$. The number of D5-branes is determined by the value of $\cal{R}(\eta)$ at the points $\eta=1,2,3,$ etc. With this we get a total of $NP(P-1)/2$ D5-branes. Finally, we also have  $P$ NS5-branes.  In fig. \ref{FIGTNP} we give this quiver\footnote{Fig. adapted from Ref. \cite{Legramandi:2021uds}}.

The potential can be computed by replacing (\ref{rankTNP}) in Eq. (\ref{solutionVhat}) and is given by 
\begin{equation}
\hat{V}(\sigma,\eta)=\sum_{k=1}^{\infty}(-1)^{k+1}\frac{NP^{3}}{k^{3}\pi^{3}}\sin\left(\frac{k\pi}{P}\right)\sin\left(\frac{k\pi}{P}\eta\right)e^{-\frac{k\pi}{P}|\sigma|}.
\end{equation}
Since the theory is trustable only for large $P$, we must expand the potential. The expression was found in Ref. \cite{Legramandi:2021uds} and is given by
\begin{equation}\label{VTNP}
\hat{V}=\frac{iNP^{2}}{2\pi^{2}}\left(-\text{Li}_{2}\left(-e^{-\frac{\pi(i\eta+\sigma)}{P}}\right)-\text{Li}_{2}\left(-e^{-\frac{\pi(\sigma-i\eta)}{P}}\right)\right).
\end{equation}

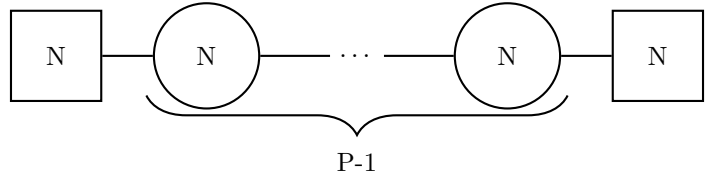
\begin{figure}[h!]
\begin{center}
	\begin{tikzpicture}
	\node (1) at (-4,0) [rectangle,draw,thick,minimum size=1.2cm] {N};
	\node (2) at (-2,0) [circle,draw,thick,minimum size=1.4cm] {N};
	\node (3) at (0,0)  {$\dots$};
	\node (5) at (4,0) [rectangle,draw,thick,minimum size=1.2cm] {N};
	\node (4) at (2,0) [circle,draw,thick,minimum size=1.4cm] {N};
	\draw[thick] (1) -- (2) -- (3) -- (4) -- (5);
	\draw [decorate,decoration={brace,amplitude=15pt,mirror},thick,yshift=-1.5em]
	(-2.8,0) -- (2.8,0) node[midway,yshift=-2.5em]{P-1};
	\end{tikzpicture}
	\caption{The $+_{P,N}$ quiver with $N$ D7-branes at $\eta = 1$, $N$ D7-branes at $\eta= P-1$ and $(P-1)N$ D5-branes.}
	\label{FIG+PN}
\end{center}

\end{figure}

For $+_{P,N}$ theory  the  associated rank function is defined as
\begin{equation}\label{rank+PN}
 {\cal R}(\eta) = \begin{cases} 
N\eta & 0\leq \eta \leq 1 \\
N & 1\leq \eta\leq (P-1)\\
N (P-\eta) & (P-1)\leq \eta\leq P .
\end{cases}
\end{equation}
In this case we have $N$ D7-branes at $\eta = 1$, $N$ D7-branes at $\eta= P-1$.  We also have $N$ D5-branes at each point $\eta=1,2,...,P-1$. With this we get a total of $N(P-1)$ D5-branes. In  fig. \ref{FIG+PN} we give this quiver\footnote{Fig. adapted from Ref. \cite{Legramandi:2021uds} }. 

The potential can be found by replacing (\ref{rank+PN}) in Eq. (\ref{solutionVhat}) and is given by
\begin{equation}
\hat{V}(\sigma,\eta)=\sum_{k=1}^{\infty}\frac{NP^{2}}{k^{3}\pi^{3}}\sin\left(\frac{k\pi}{P}\right)\left(1+(-1)^{k+1}\right)\sin\left(\frac{k\pi}{P}\eta\right)e^{-\frac{k\pi}{P}|\sigma|}.
\end{equation}
We should point that, as before,  the above theories are trustable only in the large $P$ limit and we will see that this will give us the very simple potential \cite{Legramandi:2021uds}
\begin{equation}\label{V+PN}
\hat{V}=\frac{iNP}{2\pi^{2}}\left(\text{Li}_{2}\left(-e^{-\frac{\pi(\sigma-i\eta)}{P}}\right)-\text{Li}_{2}\left(-e^{-\frac{\pi(i\eta+\sigma)}{P}}\right)-\text{Li}_{2}\left(-e^{-\frac{\pi(\sigma-i\eta)}{P}}\right)+\text{Li}_{2}\left(-e^{-\frac{\pi(i\eta+\sigma)}{P}}\right)\right).
\end{equation}

In the next section we study the dynamics of strings in these backgrounds in order to seek for integrability in an analytic way.
\section{Dynamics of Strings in $AdS_{6}\times S^{2}\times\Sigma$.}
\label{simplesolutions}
Now we consider the dynamics of strings in the background (\ref{background}) that is the subject of this work. The string action is given by
\begin{equation}
S_{P}=\frac{1}{4\pi\alpha'}\int d^{2}\sigma\left(G_{\mu\nu}\eta^{\alpha\beta}+B_{\mu\nu}\epsilon^{\alpha\beta}\right)\partial_{\alpha}X^{\mu}\partial_{\beta}X^{\nu},\label{Action}
\end{equation}
supplemented by the Virasoro constraints
\begin{align}
T_{\tilde{\sigma}\tau} & =G_{\mu\nu}\dot{X}^{\mu}X^{\prime\nu}\approx0,\nonumber \\
T_{\tilde{\sigma}\tilde{\sigma}}=T_{\tau\tau} & =G_{\mu\nu}(\dot{X}^{\mu}\dot{X}^{\nu}+X^{\prime\mu}X^{\prime\nu})\approx0.\label{VirasoroGeneral}
\end{align}

Our soliton is a string at the center of the $AdS$ space, which
rotates and wraps on the following coordinates ($\tau$ and $\tilde{\sigma}$
are the world-sheet coordinates)
\begin{equation}
t=t(\tau),\,\eta=\eta(\tau),\,\sigma=\sigma(\tau),\,\chi=\chi(\tau),\,\xi=\kappa\tilde{\sigma}.\label{StringConfiguration}
\end{equation}
Here $\kappa$ is an integer number that indicates how many times the string wraps the corresponding direction. It is straightforward to check that this truncation is consistent, since they are a solution of the generic equations of motion. For this, the coordinates above must be subject to the equations of motion as described below. We get the effective Lagrangian
\begin{equation}
L=f_{1}\dot{t}^{2}+f_{1}f_{2}(\kappa^{2}\sin^{2}(\chi)-\dot{\chi}^{2})-f_{1}f_{3}(\dot{\sigma}^{2}+\dot{\eta}^{2})+2f_{4}\dot{\chi}\kappa\sin\chi\label{Lagrangian}
\end{equation}
and 
\begin{equation}
T_{\tilde{\sigma}\tilde{\sigma}}=T_{\tau\tau}=-f_{1}\dot{t}^{2}+f_{1}f_{2}(\kappa^{2}\sin^{2}(\chi)+\dot{\chi}^{2})+f_{1}f_{3}(\dot{\sigma}^{2}+\dot{\eta}^{2})=0,\quad T_{\tilde{\sigma}\tau}=0.\label{Constraint}
\end{equation}

The equations of motion can be obtained from the Lagrangian above
and are given by

\begin{align}
f_{1}\dot{t} & =E,
\label{EoM t}
\\
f_{1}f_{2}\ddot{\chi} & =-\dot{\chi}[\dot{\sigma}\partial_{\sigma}+\dot{\eta}\partial_{\eta}](f_{1}f_{2})+\kappa\sin\chi[\dot{\sigma}\partial_{\sigma}+\dot{\eta}\partial_{\eta}]f_{4}-\kappa^{2}f_{1}f_{2}\sin(\chi)\cos(\chi),\label{EoM chi}
\\
f_{1}f_{3}\ddot{\sigma} & =-\dot{\sigma}\dot{\eta}\partial_{\eta}(f_{1}f_{3})-\frac{1}{2}\frac{E^{2}}{f_{1}}\partial_{\sigma}\log f_{1}+\frac{1}{2}\partial_{\sigma}(f_{1}f_{3})(\dot{\eta}^{2}-\dot{\sigma}^{2})\nonumber \\
 & -\frac{1}{2}\partial_{\sigma}(f_{1}f_{2})(\kappa^{2}\sin^{2}(\chi)-\dot{\chi}^{2})-\partial_{\sigma}f_{4}\dot{\chi}\kappa\sin\chi,
 \label{EoM sigma}
 \\
f_{1}f_{3}\ddot{\eta} & =-\dot{\sigma}\dot{\eta}\partial_{\sigma}(f_{1}f_{3})-\frac{1}{2}\frac{E^{2}}{f_{1}}\partial_{\eta}\log f_{1}+\frac{1}{2}\partial_{\eta}(f_{1}f_{3})(\dot{\sigma}^{2}-\dot{\eta}^{2})\nonumber
\\
 & -\frac{1}{2}\partial_{\eta}(f_{1}f_{2})(\kappa^{2}\sin^{2}(\chi)-\dot{\chi}^{2})-\partial_{\eta}f_{4}\dot{\chi}\kappa\sin\chi.
 \label{EoM eta}
\end{align}
In the first of the equations above, $E$ is a constant of integration
and has been used in the last three equations. It is easy to verify that the derivative of the Virasoro constraints (\ref{Constraint}) vanishes if Eqs. (\ref{EoM t}-\ref{EoM eta}) are used. Since Eqs. (\ref{EoM t}-\ref{EoM eta}) define the $\tau$ evolution of the string
configuration, we can study its (non) integrability. In the next, we consider the possibility 
of finding simple solutions and study these aspects for the configuration (\ref{StringConfiguration}).

\subsection{Finding Simple Solutions.}

The first step we take is to look for some simple solutions of the eom (\ref{EoM t}-\ref{EoM eta}). As cited in the introduction, the general procedure is to find a solution to Eqs. (\ref{EoM sigma}) and  (\ref{EoM eta}) which must be replaced in the NVE of Eq. (\ref{EoM chi}). However, we just need to solve (\ref{EoM sigma}) or (\ref{EoM eta}) and use the constraint (\ref{Constraint}). We will see that, with this at hand, we obtain general conclusions without choosing any specific form of the background. In the next sections we will apply this idea to many cases. First we note that 
\[
\ddot{\chi}=\dot{\chi}=\chi=0
\]
is a solution to the second equation in (\ref{EoM chi}). Replacing
this in the other equations we get
\begin{align*}
\ddot{\sigma} & =-\dot{\sigma}\dot{\eta}\partial_{\eta}\ln(f_{1}f_{3})-\frac{1}{2}\frac{E^{2}}{f_{1}^{2}f_{3}}\partial_{\sigma}\ln f_{1}+\frac{1}{2}\partial_{\sigma}\ln(f_{1}f_{3})(\dot{\eta}^{2}-\dot{\sigma}^{2}),\\
\ddot{\eta} & =-\dot{\sigma}\dot{\eta}\partial_{\sigma}\ln(f_{1}f_{3})-\frac{1}{2}\frac{E^{2}}{f_{1}^{2}f_{3}}\partial_{\eta}\ln f_{1}+\frac{1}{2}\partial_{\eta}\ln(f_{1}f_{3})(\dot{\sigma}^{2}-\dot{\eta}^{2}).
\end{align*}

The equations above can be further simplified. By using equation (\ref{EoM t}), the constraint can be written
as 
\begin{equation}
\dot{\sigma}^{2}+\dot{\eta}^{2}=\frac{E^{2}}{f_{1}^{2}f_{3}}.\label{FinalConstraint}
\end{equation}
With this we get
\begin{equation}
\ddot{\sigma}=-\dot{\sigma}\dot{\eta}\partial_{\eta}\ln(f_{1}f_{3})+\frac{1}{2}\frac{E^{2}}{f_{1}^{2}f_{3}^{2}}\partial_{\sigma}f_{3}-\dot{\sigma}^{2}\partial_{\sigma}\ln(f_{1}f_{3}),\label{SigmaEq}
\end{equation}
and 
\begin{equation}
\ddot{\eta}=-\dot{\sigma}\dot{\eta}\partial_{\sigma}\ln(f_{1}f_{3})+\frac{1}{2}\frac{E^{2}}{f_{1}^{2}f_{3}^{2}}\partial_{\eta}f_{3}-\dot{\eta}^{2}\partial_{\eta}\ln(f_{1}f_{3}).\label{EtaEq}
\end{equation}

Finally, we fluctuate $\chi$ by $\chi=0+z(\tau)$
in equation (\ref{EoM chi}) to get the NVE 
\begin{equation}
\frac{d^2z(\tau)}{d\tau^2} +{\cal B}\frac{dz(\tau)}{d\tau}+{\cal A}z(\tau)=0, \nonumber
\end{equation}
\begin{equation}
{\cal B}=[\dot{\sigma}\partial_{\sigma}+\dot{\eta}\partial_{\eta}]\ln(f_{1}f_{2}), \nonumber
\end{equation}
\begin{equation}
\:{\cal A}=\kappa^{2}-\frac{\kappa}{f_{1}f_{2}}[\dot{\sigma}\partial_{\sigma}+\dot{\eta}\partial_{\eta}]f_{4}.\label{NVE}
\end{equation}
The coefficients $\cal{A}$ and $\cal{B}$ depend on $\sigma,\eta$. Therefore, in principle, we should solve for $\sigma,\eta$ in such a way that the NVE becomes a linear second order differential equation. However, if we choose the simple solution $\sigma=\sigma_{0}=constant$ or $\eta=\eta_{0}=constant$, we see that a linear equation can be obtained. We analyse now both cases.

\subsection{The Case $\sigma=\sigma_{0}$.}
We note that we can have a simple solution of Eq. (\ref{SigmaEq}) given by
\begin{equation}
\sigma=\sigma_{0}\mbox{ if }\frac{1}{f_{1}^{2}f_{3}^{2}}\partial_{\sigma}f_{3}|_{\sigma=\sigma_{0}}=0.\label{sigmconst}
\end{equation}
With this the NVE (\ref{NVE}) and $\eta$ equations are simplified to 
\begin{equation}
\ddot{z}+{\cal B}\dot{z}+{\cal A}z=0,\,{\cal B}=\dot{\eta}\partial_{\eta}\ln(f_{1}f_{2}),\,\:{\cal A}=\kappa^{2}-\frac{\kappa}{f_{1}f_{2}}\dot{\eta}\partial_{\eta}f_{4},
\end{equation}
and
\begin{equation}
\ddot{\eta}=\frac{1}{2}\frac{E^{2}}{f_{1}^{2}f_{3}^{2}}\partial_{\eta}f_{3}-\dot{\eta}^{2}\partial_{\eta}\ln(f_{1}f_{3}).
\label{EtaEqsigma0}
\end{equation}
Despite the simplification, the coefficients of the NVE are yet $\eta$ dependent. The general procedure is to choose a specific background and solve for $\eta$. However, depending on the background, solve equation (\ref{EtaEqsigma0}) and determine $\eta$ can be a very difficult task. Besides this, we would need to choose a specific background, which would spoil the generality of this study. 

In order to solve that, we remember that the constraint (\ref{FinalConstraint}) reduces to
\begin{equation}
\dot{\eta}^{2}=\frac{E^{2}}{f_{1}^{2}f_{3}}|_{\sigma=\sigma_{0}}.\label{Etacontraint}
\end{equation}
Now, by using the result above in Eq. (\ref{EtaEqsigma0}) we see that the equation of motion for $\eta$ can be written as
\begin{equation}
\ddot{\eta}=-\frac{E^{2}}{f_{1}^{2}f_{3}}\partial_{\eta}\ln f_{1}\sqrt{f_{3}}|_{\sigma=\sigma_{0}}.\label{etaeqconstsigma}
\end{equation}
From equations (\ref{Etacontraint}) and (\ref{etaeqconstsigma})
we see that $\dot{\eta}$ and $\ddot{\eta}$ depends only on $\eta(\tau)$.
This suggests that we use $\tau=\tau(\eta)$ and we arrive at a new NVE given by
\[
z''+{\cal D}z'+{\cal C}z=0,\,{\cal D}=(\frac{\ddot{\eta}}{\dot{\eta}^{2}}+\frac{{\cal B}}{\dot{\eta}}),\,{\cal C}=\frac{1}{\dot{\eta}^{2}}\left(\kappa^{2}-\frac{\kappa}{f_{1}f_{2}}\dot{\eta}\partial_{\eta}f_{4}\right).
\]
Now we use (\ref{Etacontraint}) and (\ref{etaeqconstsigma}) to get
\begin{equation}
z''+{\cal D}z'+{\cal C}z=0,\,{\cal D}=\partial_{\eta}\ln\left(\frac{f_{2}}{\sqrt{f_{3}}}\right),\,{\cal C}=(\frac{\kappa}{E})^{2}f_{1}^{2}f_{3}-\frac{\kappa}{E}\frac{\sqrt{f_{3}}}{f_{2}}\partial_{\eta}f_{4},\label{sigmaconst}
\end{equation}
where the quantities above must be taken at $\sigma=\sigma_{0}$ and $z'$ means derivative with respect to $\eta$. Therefore, there is no need to solve the equation for $\eta$ in order to obtain the final NVE.

\subsection{The Case $\eta=\eta_{0}$.}

In the next step, we consider that a simple solution of Eq. (\ref{EtaEq}) can be found
and is given by
\begin{equation}
\eta=\eta_{0}\mbox{ if }\frac{1}{f_{1}^{2}f_{3}^{2}}\partial_{\eta}f_{3}|_{\eta=\eta_{0}}=0.\label{etconst}
\end{equation}
In this case the constraint (\ref{FinalConstraint}) simplifies and
we get 
\begin{equation}
\dot{\sigma}^{2}=\frac{E^{2}}{f_{1}^{2}f_{3}}|_{\eta=\eta_{0}}.\label{Sigmacontraint}
\end{equation}
From the equations above we could determine $\sigma$. However, as
in the $\sigma=\sigma_{0}$ case this will not be necessary. By using
(\ref{etconst}) and (\ref{Sigmacontraint}), the $\sigma$ equation
(\ref{SigmaEq}) becomes
\begin{equation}
\ddot{\sigma}=-\frac{E^{2}}{f_{1}^{2}f_{3}}\partial_{\sigma}\ln(f_{1}\sqrt{f_{3}})|_{\eta=\eta_{0}}.\label{sigmaeqconsteta}
\end{equation}

From equations (\ref{Sigmacontraint}) and (\ref{sigmaeqconsteta})
we see that $\dot{\sigma}$ and $\ddot{\sigma}$ depend only on $\sigma(\tau)$.
This suggests that we use the parameter $\tau=\tau(\sigma)$ and the
NVE, equation (\ref{NVE}), becomes 
\[
z''+{\cal D}z'+{\cal C}z=0,\,{\cal D}=(\frac{\ddot{\sigma}}{\dot{\sigma}^{2}}+\frac{{\cal B}}{\dot{\sigma}}),\,{\cal C}=\frac{1}{\dot{\sigma}^{2}}\left(\kappa^{2}-\frac{\kappa}{f_{1}f_{2}}\dot{\sigma}\partial_{\sigma}f_{4}\right).
\]
Now, by using (\ref{Sigmacontraint}) and (\ref{sigmaeqconsteta})
we finally write
\begin{equation} \label{etaconst}
z''+{\cal D}z'+{\cal C}z=0,\,{\cal D}=\partial_{\sigma}\ln\left(\frac{f_{2}}{\sqrt{f_{3}}}\right),\,{\cal C}=(\frac{\kappa}{E})^{2}f_{1}^{2}f_{3}-\frac{\kappa}{E}\frac{\sqrt{f_{3}}}{f_{2}}\partial_{\sigma}f_{4}
\end{equation}
where the prime means a derivative with $\sigma$ and the quantities above  must by taken at $\eta=\eta_{0}$. Again, we point that there is no need to solve the $\sigma$ equation in order to obtain the second order linear differential equation. 

The results of the last subsections show that the behaviour of 
\[
\frac{1}{f_{1}^{2}f_{3}^{2}}\partial_{\eta}f_{3},\,\frac{1}{f_{1}^{2}f_{3}^{2}}\partial_{\sigma}f_{3}
\]
is crucial in order to discover what is the consistent truncation of our system of equations. In the next sections we will apply the results above and analyse some specific
backgrounds. Later we generalize our results. 

\subsection{The Kovacic's Criteria of Liouvillian Integrability.}

As shown in the last subsections, we can find consistent truncations of string equations in order to get the NVE as a homogeneous second order linear equation. With this at hand, we can study Liouvillian integrability. Interestingly, Kovacic provided not only an algorithm to find the solutions, but also a set of necessary, but not sufficient, conditions to analyse if an equation is Liouvillian integrable. Consider the general NVE in Eq. (\ref{NVE}). First, we transform it to a first order differential equation in the following way
\begin{equation}
 z(x)= e^{\int (y(x)-\frac{{\cal B}(x)}{2}) dx}\Rightarrow y'(x)-y(x)^{2}=U(x),
\end{equation}
where 
\begin{equation}\label{U-function}
4U(x)=2{\cal B}'+{\cal B}^{2}-4{\cal A}.
\end{equation}

With the potential  $U(x)$, Kovacic set some general conditions for integrability. First of all, his criteria is valid only if $U(x)$ is a fractional polynomial. If this is the case, the conditions are the following: 

\begin{itemize}
\item  Case 1: every pole of $U(x)$ has order 1 or has even
order. The order of the function $U(x)$ at infinity is either even
or greater than 2. 

\item Case 2: $U(x)$ has either one pole of order 2, or poles
of odd-order greater than 2 . 

\item Case 3: the order of the poles of $U(x)$ does not exceed
2, and its order at infinity is at least 2. 
\end{itemize}

If none of the conditions above are satisfied, the analytic solution (if it exists),
is non-Liouvillian. With the  help of the criteria above, we will study integrability of the string dynamic equations in the rest of our manuscript. The interested reader can find more detailed explanations and specific examples of this method in Refs.\cite{Nunez:2018ags,Nunez:2018qcj,Filippas:2019puw}.

\section{The Case $\sigma=\sigma_0$.}
In this section study the possibility of obtaining a consistent truncation given by $\sigma=\sigma_0$. We apply the results of the last section to some simple specific backgrounds and, later, we consider the general case.  We first consider the T-duals, the $(p,q)$-5-brane system and finally we analyse the possibility $\sigma_0=0$ in a general way.

\subsection{Type IIA Abelian T-dual.}
We now study the abelian T-dual of the  D4/D8 system in massive Type IIA theory \cite{Lozano:2018pcp}. In the electrostatic description the potential is given by Eq. (\ref{VATD})
\begin{equation}
V_{ATD}=\frac{b_{1}}{\sigma}-b_{4}(3\eta^{2}-\sigma^{2}),\,b_{1}=\frac{81}{512},\,b_{4}=\frac{m}{486},
\end{equation}
and therefore we have
\begin{equation}
\frac{1}{f_{1}^{2}f_{3}^{2}}\partial_{\sigma}f_{3}=\frac{16 b_4 \sigma ^2}{9 \pi ^2 b_1}+\frac{4}{9 \pi ^2 \sigma }.\label{Df3ATD}
\end{equation}
As said before, in order that $\sigma=\sigma_{0}$ be a solution
of Eq. (\ref{SigmaEq}) we must obey
\[
\frac{16 b_4 \sigma ^2}{9 \pi ^2 b_1}+\frac{4}{9 \pi ^2 \sigma }=0
\]
for some $\sigma_{0}$. From the last expression we see that the boundaries $\sigma_{0}=0,\sigma_{0}=\infty$
are not solutions to the equation above. However, we can try $b_{1}+4b_{4}\sigma^{3}=0$.
Using this in Eq.( \ref{sigmaconst}) we get
\[
{\cal C}=\frac{3 \pi ^2}{2}-\sqrt{3} \pi  \sqrt{-\frac{1}{a^{2/3}}},\:{\cal D}=0,
\]
where we have used $a\equiv b_1/(4b_4)$. Since $a>0$, we get complex coefficients and $\sigma=\sigma_0$ is not a consistent truncation of the equations of motion.

\subsection{Type IIA non-Abelian T-dual.}
The next simple background is given by the non-Abelian T-dual of the type IIA D4/D8 system. The electrostatic description is given by (\ref{VNATD}), namely 
\begin{equation}
V_{NATD}=\frac{a_{1}\eta}{\sigma}+4a_{4}(\eta\sigma^{2}-\eta^{3}),a_{1}=\frac{1}{128},a_{4}=\frac{m}{432},
\end{equation}
and 
\begin{equation}
\frac{1}{f_{1}^{2}f_{3}^{2}}\partial_{\sigma}f_{3}=\frac{64 a_4 \sigma ^2}{9 \pi ^2 a_1}+\frac{4}{9 \pi ^2 \sigma }.\label{Df3NATD}
\end{equation}
For $\sigma=\sigma_0$ to be a solution we need that 
$$
\frac{64 a_4 \sigma ^2}{9 \pi ^2 a_1}+\frac{4}{9 \pi ^2 \sigma }=0
$$
Just as with the abelian case, $\sigma=0$ and $\sigma=\infty$ are not solutions. However, we can try $16a_{4}\sigma^{3}+a_{1}=0$. In this case we get, from Eq. ( \ref{sigmaconst}), that
$$
{\cal C}=\frac{2\pi\sqrt{3}\sqrt{-a^{2/3}}}{2\eta^{2}-a^{2/3}}-\pi\sqrt{3}\sqrt{-a^{-2/3}}+\frac{3\pi^{2}}{2},{\cal D}=\frac{2}{\eta}-\frac{4\eta}{2\eta^{2}-a^{2/3}}.,
$$
where we have used $a\equiv a_1/(16a_4)$. Since $a>0$ we get complex coefficients  $\sigma=\sigma_{0}$ is not a consistent truncation of the eom.

\subsection{The $\tilde{T}_{N,P}$ and $+_{P,N}$ Theories.}

As said before, the theory above is trustable only for large $P$. The $\tilde{T}_{N,P}$ solution is studied in Ref.\cite{Legramandi:2021uds} with potential given by Eq. (\ref{VTNP}). Now we expand it around $\sigma=0$ to get
\[
\hat{V}=\frac{\eta(NP\log(2))}{\pi}-\frac{\pi\left(\eta^{2}+\eta\right)N}{24P}-\frac{1}{2}\sigma(\eta N)+\frac{\pi\eta N\sigma^{2}}{4P}.
\]
In Ref.\cite{Roychowdhury:2021jqt} the author gives this potential up to order $\sigma$.
However we introduced terms of order $\sigma^{2}$ since the background
functions depends on second derivative of $\sigma$. With the potential above we get 
\[
\frac{1}{(f_{1}f_{3})^{2}}\partial_{\sigma}f_{3}=\frac{16\eta\sigma}{-3\pi^{2}\eta^{2}-18\pi^{2}\eta\sigma^{2}-3\pi^{2}\eta+72\eta P^{2}\log(2)+2\pi^{2}\sigma^{2}}
\]
and therefore $\sigma=0$ is a good truncation. With this we get
\[
{\cal C}=\frac{9\pi^{2}}{4}-\frac{5\left(\pi^{2}\right)}{4P\sqrt{\log(2)}}\frac{1}{\sqrt{\eta}},{\cal D}=\frac{2}{\eta}-\frac{4\left(5\pi^{4}\eta^{3}-12\pi^{2}\eta P^{2}\right)}{5\pi^{4}\eta^{4}-24\pi^{2}\eta^{2}P^{2}+12P^{4}}+\frac{3\pi}{2\left(\pi\eta-\sqrt{6}P\right)}+\frac{3\pi}{2\left(\pi\eta+\sqrt{6}P\right)}.
\]
Then the potential becomes 
\[
U(\eta)=-9\pi^{2}+\frac{9}{4\eta^{2}}+\frac{5\pi^{2}}{P\sqrt{\log(2)}}\frac{1}{\sqrt{\eta}}-\frac{7\pi^{2}}{16P^{2}(\eta\log(2))}.
\]
This last potential is not polynomial and we cannot use the Kovacic's
criteria. However we can change the $\eta$ variable to $y=\sqrt{\eta}$, compute a new NVE and write the following $\tilde{U}(y)$:
\begin{equation}
\tilde{U}(y)=\frac{3}{4 y^2}-9 \pi ^2 y^2-\frac{3 \sqrt{\frac{3}{2}} P \pi }{2 \left(-\sqrt{6} P+\pi  y^2\right)^2}-\frac{39 \pi }{2 \left(-\sqrt{6}
   P+\pi  y^2\right)}+\frac{3 \sqrt{\frac{3}{2}} P \pi }{2 \left(\sqrt{6} P+\pi  y^2\right)^2} 
\end{equation}
\begin{equation}
   -\frac{39 \pi }{2 \left(\sqrt{6} P+\pi 
   y^2\right)} +\frac{4032 P^4 \pi ^4 y^6}{\left(12 P^4-24 P^2 \pi ^2 y^4+5 \pi ^4 y^8\right)^2}+\frac{10 \left(18 P^2 \pi ^2 y^2+19 \pi
   ^4 y^6\right)}{12 P^4-24 P^2 \pi ^2 y^4+5 \pi ^4 y^8}+\frac{5 \pi  y}{4 \sqrt{\log (2)}}.\nonumber
\end{equation}
Now we have it in a polynomial form and, even after that procedure, this last potential does not obey any of the Kovacic's criteria: we conclude that this model is not integrable.

Next we consider the $+_{P,N}$ theory. This solution, studied  in Ref.\cite{Legramandi:2021uds}, has potential given by Eq. (\ref{V+PN}). By expanding it in $\sigma$ we get
\[
\hat{V}=\frac{N}{4\pi}\left(-2\text{\ensuremath{\eta}Log}\left(1-\eta^{2}\right)-\left(\eta^{2}+1\right)\log\left(\frac{\eta+1}{1-\eta}\right)+\eta(6+4\log(2))-4\eta\log\left(\frac{\pi}{P}\right)\right).
\]
From the result above, we have 
\[
\frac{1}{(f_{1}f_{3})^{2}}\partial_{\sigma}f_{3}=\sigma h(\eta)
\]
with
\begin{align*}
h(\eta) & =-\frac{8\log\left(\frac{\eta+1}{1-\eta}\right)}{9\pi^{2}\left(\eta^{2}\log\left(\frac{\eta+1}{1-\eta}\right)+2\eta\log\left(\frac{1}{4}\left(1-\eta^{2}\right)\right)-6\eta+\log\left(\frac{\eta+1}{1-\eta}\right)+4\eta\log\left(\frac{\pi}{P}\right)\right)}\\
 & -\frac{16\eta}{9\pi^{2}\left(\eta^{2}-1\right)^{2}\log\left(\frac{\eta+1}{1-\eta}\right)}+{\cal O}(\sigma^{3}).
\end{align*}
Hence the truncation $\sigma=0$ is consistent. We can also obtain
the U function. Nevertheless, we see that we get a logarithm dependence in $\eta$. This potential,
thus, cannot be written as a polynomial function and we cannot apply the Kovacic's criteria. 

We point that the cases above were studied in Ref.  \cite{Roychowdhury:2021jqt}. The
author used a further truncation $\eta=0$. However, if we use  $\eta=\sigma=\chi=0$ in (\ref{Constraint}), we find $\dot{t}f_1=0$. Then we get $\dot{t}=0$ or $f_1=0$. The first choice is not consistent and the second give us a null background making the equations of motion trivially null. Therefore, with this choice it is not possible to conclude about integrability. From the viewpoint given in our work there is no need to fix $\eta=0$ to make the truncation consistent. Despite this, due to the shape of the potential, the Kovacic's criteria were applied only for the $\tilde{T}_{N,P}$.

\section{The Case $\eta=\eta_0$.}
In the previous section we learnt that $\sigma=\sigma_0$ sometimes is a good solution for the string's equations of motion, sometimes not. In this section we apply the results of section \ref{simplesolutions} to search the other possibility of obtaining a simple solution for $\eta=\eta_{0}$. We  show that this works for several backgrounds: the abelian and non-abelian T-duals, the T$_{N}$, +$_{MN}$, $\tilde{T}_{N,P}$ and $+_{P,N}$ theories. Finally, we consider the general case, which includes any long quiver.

\subsection{Type IIA Abelian T-dual}
As we already know, in this case the potential is given by Eq. (\ref{VATD})
and therefore we have
\begin{equation}
\frac{1}{f_{1}^{2}f_{3}^{2}}\partial_{\eta}f_{3}=0.\label{Df3ATDeta}
\end{equation}
We should point out that the equation above is valid for any value of $\eta$.
Then, the coefficients of the NVE (\ref{etaconst}) becomes
\[
{\cal C}=-\frac{9 \pi ^2 b_1}{4 \left(2 b_4 \sigma ^3-b_1\right)},\:{\cal D}=\frac{9 b_1}{2 \sigma  \left(2 b_4 \sigma ^3-b_1\right)}+\frac{4}{\sigma },
\]
where we have used $\kappa=E$. The coefficients
are already rational functions and can be analysed with the Kovacic's
criteria. We get a potential (\ref{U-function}) given by
\begin{eqnarray*}
4U&=&\frac{9a^{2/3}}{4\left(a^{2/3}+\sqrt[3]{a}\sigma+\sigma^{2}\right)^{2}}-\frac{3\left(1+\pi^{2}a^{2/3}\right)\sigma}{\sqrt[3]{a}\left(a^{2/3}+\sqrt[3]{a}\sigma+\sigma^{2}\right)}+\frac{3\left(1+4\pi^{2}a^{2/3}\right)}{2\left(a^{2/3}+\sqrt[3]{a}\sigma+\sigma^{2}\right)}
\\
&&+\frac{3\left(\pi^{2}a^{2/3}+1\right)}{\sqrt[3]{a}\left(\sigma-\sqrt[3]{a}\right)}-\frac{3}{4\left(\sigma-\sqrt[3]{a}\right)^{2}}+\frac{5}{4\sigma^{2}}.
\end{eqnarray*}
In the expression above we have used $b_1=2ab_4$ in order to simplify it. By analysing the $U-$function, we see that it does not satisfy all the three possible necessary Kovacic's conditions described above. The solution to the equation should then be non-Liouvillian. Then $\eta=\eta_{0}$ give us a consistent truncation in order to study integrability.     

\subsection{Type IIA non-Abelian T-dual}
For this case we have the potential given by Eq. (\ref{VNATD}) and 
\begin{equation}
\frac{1}{f_{1}^{2}f_{3}^{2}}\partial_{\eta}f_{3}=0.\label{Df3NATDeta}
\end{equation}
As in the abelian case, again we have the equation above valid for any value
of $\eta$ and a simple solution to (\ref{EtaEq}) is given by $\eta=\eta_{0}$.
Therefore, the coefficients of the NVE ((\ref{etaconst})) become
$$
{\cal C}  =-\frac{9 \pi  a^2 \eta_{0}  \sqrt{\frac{\sigma }{a-\sigma ^3}}}{2 \sigma  \left(a^2+9 a \eta_{0} ^2 \sigma -2 a \sigma ^3+\sigma ^6\right)}-\frac{45 \pi  a \eta_{0}  \sigma ^2 \sqrt{\frac{\sigma }{a-\sigma ^3}}}{2 \left(a^2+9 a \eta_{0} ^2 \sigma -2 a \sigma ^3+\sigma ^6\right)}+\frac{9 \pi ^2 a}{4 \left(a-\sigma ^3\right)},
$$
and
\[
{\cal D}=-\frac{3 \left(3 a \eta_{0} ^2-2 a \sigma ^2+2 \sigma ^5\right)}{a^2+9 a \eta_{0} ^2 \sigma -2 a \sigma ^3+\sigma ^6}+\frac{9 \sigma ^2}{2 \left(\sigma ^3-a\right)}+\frac{1}{2 \sigma },
\]
where we have used $\kappa=E$ and $a=a_{1}/(8a_{4})$. We see that the coefficients above are not rational functions. We could try to solve this with a change of coordinate but, since the term in the square root is cubic, it is not possible to find an inverse. But we can note that all terms with square roots cancel at $\eta_0=0$. Therefore we choose this particular value to get 
$$
{\cal C}=-\frac{9 \pi ^2 a}{4 \left(\sigma ^3-a\right)},\;{\cal D}=\frac{1}{2 \sigma }-\frac{3 \sigma ^2}{2 \left(\sigma ^3-a\right)}.
$$
The coefficients above are already rational functions and we find from Eq.(\ref{U-function}) that 
$$
4U =-\frac{3}{4 \sigma ^2}-\frac{15 a^{2/3}}{4 \left(a^{2/3}+\sqrt[3]{a} \sigma +\sigma ^2\right)^2}+\frac{-12 \pi ^2 a^{2/3}-6 \pi ^2 \sqrt[3]{a} \sigma +5}{2 \left(a^{2/3}+\sqrt[3]{a} \sigma +\sigma ^2\right)}+\frac{3 \pi ^2 \sqrt[3]{a}}{\sigma -\sqrt[3]{a}}+\frac{5}{4 \left(\sigma -\sqrt[3]{a}\right)^2}
$$
Analysing the $U-$function, we see that it does not satisfy all the three possible necessary Kovacic's conditions and the solution must be non-Liouvillian. However, this case
is less general if compared to the abelian one, since we had to choose $\eta_{0}=0$. 
We can also conclude that $\eta=0$ provides a consistent solution that allow us to analyse the integrability of the system. 

\subsection{$T_{N}$ Theory}\label{secTN}
This solution is studied in Ref. \cite{Uhlemann:2019ypp}. Its electrostatic description has potential given by (\ref{VTN}) \cite{Nunez:2018ags}
\[
\hat{V}=\frac{9N^{2}}{32\pi^{2}}\sum_{k=1}^{\infty}\frac{(-1)^{k+1}}{k^{2}}\sin\left(\frac{4k\pi}{9N}\eta\right)e^{-\frac{4k\pi}{9N}|\sigma|}.
\]
Since we look for a solution with $\eta=0$, we can expand the potential above to obtain
\begin{equation}\label{expansion}
\hat{V}(\sigma,\eta)\approx\eta\rho(\sigma)-\frac{\eta^{3}}{6}\beta(\sigma)+\frac{\eta^{5}}{24}\gamma(\sigma)+{\cal O}(\eta^7),
\end{equation}
with
\[
\rho=-\frac{N}{8\pi}\sum_{k=1}^{\infty}\frac{(-1)^{k}}{k}e^{-\frac{4k\pi}{9N}|\sigma|}=-\frac{N}{8\pi}\mbox{Li}_{1}(-e^{-\frac{4\pi}{9N}|\sigma|})=\frac{N}{8\pi}\ln(1+e^{-\frac{4\pi}{9N}|\sigma|}),
\]
\[
\beta=-\frac{2\pi}{81N}\sum_{k=1}^{\infty}(-1)^{k}ke^{-\frac{4k\pi}{9N}|\sigma|}=-\frac{2\pi}{81N}\mbox{Li}_{-1}(-e^{-\frac{4\pi}{9N}|\sigma|})=\frac{2\pi}{81N}\frac{1}{(e^{\frac{2\pi}{9N}|\sigma|}+e^{-\frac{2\pi}{9N}|\sigma|})^{2}},
\]
and
\begin{eqnarray}
\gamma&=&-\frac{1}{2\sigma}\left(\frac{4\pi}{9N}\right)^{3}\sum_{k=1}^{\infty}(-1)^{k}k^{3}e^{-\frac{4k\pi}{9N}|\sigma|}=-\frac{1}{2\sigma}\left(\frac{4\pi}{9N}\right)^{3}\mbox{Li}_{-3}(-e^{-\frac{4\pi}{9N}|\sigma|})\nonumber\\
&=&\frac{1}{2\sigma}\left(\frac{4\pi}{9N}\right)^{3}\frac{e^{-\frac{4\pi}{9N}|\sigma|}(1-4e^{-\frac{4\pi}{9N}|\sigma|}+e^{-\frac{8\pi}{9N}|\sigma|})}{(1+e^{-\frac{4\pi}{9N}|\sigma|})^{4}}.
\end{eqnarray}
With the quantities above we get 
\begin{equation}\label{df1}
\frac{1}{(f_{1}f_{3})^{2}}\partial_{\eta}f_{3}=\eta\frac{A(\sigma)}{4374N^{4}B(\sigma)}+{\cal O}(\eta^2)
\end{equation}
where
\begin{eqnarray}
A(\sigma)	&=&\sigma e^{\frac{4\pi\sigma}{9N}}\left(2916\pi N^{4}\sigma-6561N^{5}-20480\pi^{5}\right)+\sigma e^{\frac{8\pi\sigma}{9N}}\left(-2916\pi N^{4}\sigma-6561N^{5}+5120\pi^{5}\right)\nonumber\\
	&&+11520\pi^{4}N\left(-3e^{\frac{4\pi\sigma}{9N}}-3e^{\frac{8\pi\sigma}{9N}}+e^{\frac{4\pi\sigma}{3N}}+1\right)\log\left(e^{-\frac{4\pi\sigma}{9N}}+1\right)
\end{eqnarray}
and	
\begin{eqnarray}\label{B}
B(\sigma)&=&24\pi\sigma+e^{\frac{8\pi\sigma}{9N}}\left(24\pi\sigma+18N\sigma^{2}\right)
\\
&&+e^{\frac{4\pi\sigma}{9N}}\left(48\pi\sigma+18N\sigma^{2}\right)+54N\left(e^{\frac{4\pi\sigma}{9N}}+1\right)^{3}\log\left(e^{-\frac{4\pi\sigma}{9N}}+1\right).\nonumber
\end{eqnarray}
With this, it is clear that in the limit  $\eta=0$, Eq. (\ref{df1}) is null. Therefore, it is a consistent truncation in order to study the NVE (\ref{etaconst}). The same kind of procedure will be repeated in all the next cases. The coefficients are given by
\begin{align*}
{\cal C} & =\frac{9 \pi ^2 \left(\sigma  \left(4 \pi  \left(e^{\frac{4 \pi  \sigma }{9 N}}+1\right)+3 N \sigma  e^{\frac{4 \pi  \sigma }{9 N}}\right)+9 N \left(e^{\frac{4 \pi  \sigma }{9 N}}+1\right)^2 \log \left(e^{-\frac{4 \pi  \sigma }{9 N}}+1\right)\right)}{4 \left(e^{\frac{4 \pi  \sigma }{9 N}}+1\right) \left(9 N \left(e^{\frac{4 \pi  \sigma }{9 N}}+1\right) \log \left(e^{-\frac{4 \pi  \sigma }{9 N}}+1\right)+4 \pi  \sigma \right)},
\end{align*}
and
\[
{\cal D}=\frac{2 \pi  \left(4 \pi  \sigma -9 N \left(e^{\frac{8 \pi  \sigma }{9 N}}-1\right) \log \left(e^{-\frac{4 \pi  \sigma }{9 N}}+1\right)\right)}{9 N \left(e^{\frac{4 \pi  \sigma }{9 N}}+1\right) \left(9 N \left(e^{\frac{4 \pi  \sigma }{9 N}}+1\right) \log \left(e^{-\frac{4 \pi  \sigma }{9 N}}+1\right)+4 \pi  \sigma \right)}.
\]
The potential (\ref{U-function}) will be given by
\begin{align*}
U & =\frac{20 \pi ^2}{81 N^2 \left(e^{\frac{4 \pi  \sigma }{9 N}}+1\right)^2}-\frac{\pi ^2 \left(729 N^2-4\right)}{81 N^2}+\frac{2187 \pi  N^3 \sigma ^2+72 \pi  N \log \left(e^{-\frac{4 \pi  \sigma }{9 N}}+1\right)-144 \pi  N-64 \pi ^2 \sigma }{324 N^2 \sigma  \left(e^{\frac{4 \pi  \sigma }{9 N}}+1\right)}
\\& 
 -\frac{\pi  \left(9 N \log \left(e^{-\frac{4 \pi  \sigma }{9 N}}+1\right)+4 \pi  \sigma \right) \left(729 N^3 \sigma ^2+24 N \log \left(e^{-\frac{4 \pi  \sigma }{9 N}}+1\right)-48 N+32 \pi  \sigma \right)}{108 N^2 \sigma  \left(9 N e^{\frac{4 \pi  \sigma }{9 N}} \log \left(e^{-\frac{4 \pi  \sigma }{9 N}}+1\right)+9 N \log \left(e^{-\frac{4 \pi  \sigma }{9 N}}+1\right)+4 \pi  \sigma \right)}
 \\& 
 +\frac{20 \pi ^2 \left(9 N \log \left(e^{-\frac{4 \pi  \sigma }{9 N}}+1\right)+4 \pi  \sigma \right)^2}{81 N^2 \left(9 N e^{\frac{4 \pi  \sigma }{9 N}} \log \left(e^{-\frac{4 \pi  \sigma }{9 N}}+1\right)+9 N \log \left(e^{-\frac{4 \pi  \sigma }{9 N}}+1\right)+4 \pi  \sigma \right)^2}
 \end{align*}

We cannot apply the Kovacic's criteria to the case above since the coefficients are not fractional polynomials. However, for very large $\sigma$ it reduces to  
\[
U=\frac{243 N^2}{16}-9 \pi ^2-\frac{27 \pi  N \sigma }{4}-\frac{2187 N^3}{16 (9 N+4 \pi  \sigma )}+\frac{20 \pi ^2}{(9 N+4 \pi  \sigma )^2}+{\cal O}(e^{-c\sigma})
\]
where ${\cal O}(e^{-c\sigma})$ means terms which decrease exponentially.  Therefore, in this region, the potential has the desired shape. Nevertheless, the order of the fractional polynomial is negative and does not satisfy all Kovacic's conditions. Since we are analysing only one particular point, this is only a hint and we cannot conclude that the solution is not integrable. We need more regions to enforce this. 

\subsection{$+_{MN}$ Theory}

This solution is studied in Ref. \cite{Uhlemann:2019ypp}, with electrostatic description given in Ref.\cite{Nunez:2018ags}. The potential is given by (\ref{V+MN})
\[
\hat{V}=\frac{9MN}{32\pi^{2}}\sum_{k=1}^{\infty}\frac{1-(-1)^{k}}{k^{2}}\sin\left(\frac{4\pi k}{9M}\eta\right)e^{-\frac{4\pi k}{9M}|\sigma|}.
\]
Now we expand in $\eta$ as in the last subsection to get
\[
\rho=\frac{N}{8\pi}\sum_{k=1}^{\infty}\frac{1-(-1)^{k}}{k}e^{-\frac{4\pi k}{9M}|\sigma|}=\frac{N}{8\pi}\left(\mbox{Li}_{1}(e^{-\frac{4\pi}{9M}|\sigma|})\right)-\frac{N}{8\pi}\mbox{Li}_{1}(-e^{-\frac{4\pi}{9N}|\sigma|})=\frac{N}{8\pi}\ln\frac{(1-e^{-\frac{4\pi}{9M}|\sigma|})}{(1+e^{-\frac{4\pi}{9M}|\sigma|})},
\]
\begin{align*}
\beta & =\frac{2\pi N}{81M^{2}}\sum_{k=1}^{\infty}(1-(-1)^{k})ke^{-\frac{4\pi k}{9M}|\sigma|}=\frac{2\pi N}{81M^{2}}\left(\mbox{Li}_{-1}(e^{-\frac{4\pi}{9N}|\sigma|})-\mbox{Li}_{-1}(-e^{-\frac{4\pi}{9N}|\sigma|})\right)=\\
 & =\frac{2\pi N}{81M^{2}}\left(\frac{1}{(e^{\frac{2\pi}{9N}|\sigma|}-e^{-\frac{2\pi}{9N}|\sigma|})^{2}}+\frac{1}{(e^{\frac{2\pi}{9N}|\sigma|}+e^{-\frac{2\pi}{9N}|\sigma|})^{2}}\right)=\frac{4\pi N}{81M^{2}}\frac{e^{\frac{4\pi}{9N}|\sigma|}+e^{-\frac{4\pi}{9N}|\sigma|}}{(e^{\frac{4\pi}{9N}|\sigma|}-e^{-\frac{4\pi}{9N}|\sigma|})^{2}},
\end{align*}
and
$$
\gamma=\frac{2\pi N}{81M^{2}}\left(\frac{4\pi}{9M}\right)^{2}\left(\mbox{Li}_{-3}(e^{-\frac{4\pi}{9N}|\sigma|})-\mbox{Li}_{-3}(-e^{-\frac{4\pi}{9N}|\sigma|})\right)=\frac{2e^{-\frac{4\pi}{9N}|\sigma|}\left(e^{-\frac{8\pi}{3N}|\sigma|}+23e^{-\frac{16\pi}{9N}|\sigma|}+23e^{-\frac{8\pi}{9N}|\sigma|}+1\right)}{\left(1-e^{-\frac{8\pi}{9N}|\sigma|}\right)^{4}}.
$$

With the expressions above we find that 
\[
\frac{1}{(f_{1}f_{3})^{2}}\partial_{\eta}f_{3}=\eta\frac{64A(\sigma)}{81B(\sigma)}+{\cal O}(\eta^2),
\]
with
\begin{eqnarray*}
A(\sigma) &&=405MN\left(-21e^{\frac{8\pi\sigma}{9N}}+21e^{\frac{16\pi\sigma}{9N}}+e^{\frac{8\pi\sigma}{3N}}-1\right)\log\left(\frac{e^{\frac{4\pi\sigma}{9N}}-1}{e^{\frac{4\pi\sigma}{9N}}+1}\right)+\\
&&-2e^{\frac{4\pi\sigma}{9N}}\left(4\pi\sigma\left((45M+1)e^{\frac{16\pi\sigma}{9N}}+(990M+6)e^{\frac{8\pi\sigma}{9N}}+45M+1\right)+9N\left(e^{\frac{16\pi\sigma}{9N}}-1\right)\right)
\end{eqnarray*}
and
$$
B(\sigma)=8\pi\sigma e^{\frac{4\pi\sigma}{9N}}\left(e^{\frac{8\pi\sigma}{9N}}-1\right)\left(4\pi N\sigma\left(e^{\frac{8\pi\sigma}{9N}}+1\right)-27M^{2}\left(e^{\frac{8\pi\sigma}{9N}}-1\right)\right)+243M^{2}N\left(e^{\frac{8\pi\sigma}{9N}}-1\right)^{3}\log\left(\frac{e^{\frac{4\pi\sigma}{9N}}-1}{e^{\frac{4\pi\sigma}{9N}}+1}\right).
$$

Again we see that the last expression is null for $\eta=0$. Therefore, this is a good truncation of our system and we can study the NVE (\ref{etaconst}). The coefficients are given by
$$
{\cal C}=\frac{\pi ^2 \left(8 \pi  \sigma  e^{\frac{4 \pi  \sigma }{9 N}} \left(4 \pi  N \sigma  \left(e^{\frac{8 \pi  \sigma }{9 N}}+1\right)-27 M^2 \left(e^{\frac{8 \pi  \sigma }{9 N}}-1\right)\right)+243 M^2 N \left(e^{\frac{8 \pi  \sigma }{9 N}}-1\right)^2 \log \left(\frac{e^{\frac{4 \pi  \sigma }{9 N}}-1}{e^{\frac{4 \pi  \sigma }{9 N}}+1}\right)\right)}{12 M^2 \left(e^{\frac{8 \pi  \sigma }{9 N}}-1\right) \left(9 N \left(e^{\frac{8 \pi  \sigma }{9 N}}-1\right) \log \left(\frac{e^{\frac{4 \pi  \sigma }{9 N}}-1}{e^{\frac{4 \pi  \sigma }{9 N}}+1}\right)-8 \pi  \sigma  e^{\frac{4 \pi  \sigma }{9 N}}\right)}
$$
and
\begin{align*}
{\cal D} & =\frac{2 \pi  \left(32 \pi  \sigma  e^{\frac{4 \pi  \sigma }{3 N}}-9 N \left(-5 e^{\frac{8 \pi  \sigma }{9 N}}+5 e^{\frac{16 \pi  \sigma }{9 N}}+e^{\frac{8 \pi  \sigma }{3 N}}-1\right) \log \left(\frac{e^{\frac{4 \pi  \sigma }{9 N}}-1}{e^{\frac{4 \pi  \sigma }{9 N}}+1}\right)\right)}{9 N \left(e^{\frac{4 \pi  \sigma }{9 N}}-1\right) \left(e^{\frac{4 \pi  \sigma }{9 N}}+1\right) \left(e^{\frac{8 \pi  \sigma }{9 N}}+1\right) \left(9 N \left(e^{\frac{8 \pi  \sigma }{9 N}}-1\right) \log \left(\frac{e^{\frac{4 \pi  \sigma }{9 N}}-1}{e^{\frac{4 \pi  \sigma }{9 N}}+1}\right)-8 \pi  \sigma  e^{\frac{4 \pi  \sigma }{9 N}}\right)}.
\end{align*}

Finally, the potential (\ref{U-function}) is given by
\begin{align*}
U & =-9\pi^{2}+\frac{64\pi^{2}\left(e^{\frac{16\pi\sigma}{9N}}+2e^{\frac{24\pi\sigma}{9N}}+2e^{\frac{8\pi\sigma}{9N}}\right)}{81N^{2}\left(e^{\frac{16\pi\sigma}{9N}}-1\right)^{2}}+\frac{3\pi^{2}N^{2}}{2M^{2}}\left(e^{\frac{4\pi\sigma}{9N}}+e^{-\frac{4\pi\sigma}{9N}}\right)\log\left(\frac{e^{\frac{4\pi\sigma}{9N}}-1}{e^{\frac{4\pi\sigma}{9N}}+1}\right)
\\
 &+ \frac{4\pi^{3}N\left(e^{\frac{8\pi\sigma}{9N}}+1\right)\sigma}{3M^{2}\left(e^{\frac{8\pi\sigma}{9N}}-1\right)}+\frac{20\pi^{2}\left(e^{\frac{8\pi\sigma}{9N}}+1\right)^{2}\log^{2}\left(\frac{e^{\frac{4\pi\sigma}{9N}}-1}{e^{\frac{4\pi\sigma}{9N}}+1}\right)}{\left(8\pi\sigma e^{\frac{4\pi\sigma}{9N}}-9N\left(e^{\frac{8\pi\sigma}{9N}}-1\right)\log\left(\frac{e^{\frac{4\pi\sigma}{9N}}-1}{e^{\frac{4\pi\sigma}{9N}}+1}\right)\right)^{2}}
 \\
 & -\frac{\pi^{2}\left(e^{\frac{8\pi\sigma}{9N}}+1\right)\left(64M^{2}e^{\frac{8\pi\sigma}{9N}}+243N^{4}\left(e^{\frac{8\pi\sigma}{9N}}-1\right)^{2}\log^{2}\left(\frac{e^{\frac{4\pi\sigma}{9N}}-1}{e^{\frac{4\pi\sigma}{9N}}+1}\right)\right)}{18M^{2}Ne^{\frac{4\pi\sigma}{9N}}\left(e^{\frac{8\pi\sigma}{9N}}-1\right)\left(9N\left(e^{\frac{8\pi\sigma}{9N}}-1\right)\log\left(\frac{e^{\frac{4\pi\sigma}{9N}}-1}{e^{\frac{4\pi\sigma}{9N}}+1}\right)-8\pi\sigma e^{\frac{4\pi\sigma}{9N}}\right)}  \\
 &
 -\frac{16\pi^{2}\left(-4e^{\frac{8\pi\sigma}{9N}}+e^{\frac{16\pi\sigma}{9N}}+1\right)\log\left(\frac{e^{\frac{4\pi\sigma}{9N}}-1}{e^{\frac{4\pi\sigma}{9N}}+1}\right)}{9N\left(e^{\frac{8\pi\sigma}{9N}}-1\right)\left(9N\left(e^{\frac{8\pi\sigma}{9N}}-1\right)\log\left(\frac{e^{\frac{4\pi\sigma}{9N}}-1}{e^{\frac{4\pi\sigma}{9N}}+1}\right)-8\pi\sigma e^{\frac{4\pi\sigma}{9N}}\right)}.
\end{align*}
Again, we have that the potential above is not a fractional polynomial. However, it is simple to show that for very large $\sigma$ we have 
\[
U=-9\pi^{2}-\frac{3\pi^{2}N^{2}}{M^{2}}+\frac{4\pi^{3}N\sigma}{3M^{2}}-\frac{40\pi^{2}}{\left(8\pi\sigma+9N\right)^{2}}+\frac{27\pi^{2}N^{3}}{M^{2}(189N+8\pi\sigma)}+{\cal O}(e^{-c\sigma})
\]
where ${\cal O}(e^{-c\sigma})$ means terms which decrease exponentially.
Therefore, in this region, we can test the Kovacic's criteria by analysing the $U-$function.  We see that it does not satisfy all the conditions. As before, this is a hint and we cannot conclude that the solution is not integrable.

\subsection{The $\tilde{T}_{N,P}$ and $+_{P,N}$ Theories.}
In this subsection we analyse the integrability of the $\tilde{T}_{N,P}$ and $+_{P,N}$ theories. These theories are trustable only in the large $P$ limit (so we must expand them in $1/P$).  The $\tilde{T}_{N,P}$ solution is studied in Ref.\cite{Legramandi:2021uds} with potential given by Eq. (\ref{VTNP}). Now we must expand it around $\eta=0$ to get
\[
\hat{V}=\frac{\eta N}{24}\left(-\frac{\pi\left(\eta^{2}-3\sigma^{2}\right)}{P}+\frac{24P\log(2)}{\pi}-12\sigma\right).
\]
With the above potential we get 
\[
\frac{1}{(f_{1}f_{3})^{2}}\partial_{\eta}f_{3}=\frac{4\eta}{9a(a+\pi\sigma)}-\frac{4\eta}{9a(\pi\sigma-a)}
\]
where we have used $a^{2}=24P^{2}\log(2)$. Therefore $\eta=0$ is
a good truncation. With this we find
\[
{\cal C}=\frac{3\pi^{4}\sigma^{2}}{4(\pi\sigma-b)(b+\pi\sigma)}-\frac{18\pi^{2}P^{2}\log(2)}{(\pi\sigma-b)(b+\pi\sigma)},{\cal D}=-\frac{\pi}{2(b+\pi\sigma)}-\frac{\pi}{2(\pi\sigma-b)}.
\]
where $b^{2}=8P^{2}\log(2)$. Therefore the potential becomes 
\[
U=\frac{12\pi^{2}b^{2}+\pi^{2}}{4b(\pi\sigma-b)}+\frac{-12\pi^{2}b^{2}-\pi^{2}}{4b(b+\pi\sigma)}+\frac{5\pi^{2}}{4(\pi\sigma-b)^{2}}+\frac{5\pi^{2}}{4(b+\pi\sigma)^{2}}-3\pi^{2}.
\]
Just like the $\sigma=0$ truncation, we find a polynomial
function and we conclude, by using the Kovacic's criteria, that it
is not integrable. 

Next we consider the $+_{P,N}$ theory. This solution, studied  in Ref.\cite{Legramandi:2021uds}, has potential given by Eq. \ref{V+PN}. By expanding the potential
around $\eta=0$ we get
\[
\hat{V}=\eta^{3}\left(-\frac{\pi N}{36P^{2}}-\frac{N}{6\left(\pi\sigma^{2}\right)}\right)+\eta\left(\frac{\pi N\sigma^{2}}{12P^{2}}+\frac{N}{\pi}\left(\log(2)-\log\left(\frac{\pi\sigma}{P}\right)\right)\right)
\]
and
\[
\frac{1}{(f_{1}f_{3})^{2}}\partial_{\eta}f_{3}=-\frac{8\eta\left(18P^{2}+\pi^{2}\sigma^{2}\right)}{9\pi^{2}\left(\pi^{2}\eta^{2}\sigma^{2}+18\eta^{2}P^{2}-48P^{2}\sigma^{2}-6P^{2}\sigma^{2}\log(64)+36P^{2}\sigma^{2}\log\left(\frac{\pi\sigma}{P}\right)+\pi^{2}\sigma^{4}\right)}.
\]
Therefore, $\eta=0$ is a good truncation. We compute 
\begin{align*}
{\cal C} & =\frac{3\pi^{2}\left(36P^{2}\log\left(\frac{\pi\sigma}{P}\right)-6P^{2}(8+\log(64))+\pi^{2}\sigma^{2}\right)}{4\left(12P^{2}\log\left(\frac{\pi\sigma}{P}\right)-6P^{2}(2+\log(4))+\pi^{2}\sigma^{2}\right)},\\
{\cal D} & =\frac{-6P^{2}-\pi^{2}\sigma^{2}}{\sigma\left(12P^{2}\log\left(\frac{\pi\sigma}{P}\right)-12P^{2}-6P^{2}\log(4)+\pi^{2}\sigma^{2}\right)}-\frac{6P^{2}}{\sigma\left(6P^{2}+\pi^{2}\sigma^{2}\right)}.
\end{align*}
For the U function we find
\begin{align*}
U & =-\frac{9\left(12\pi^{4}P^{2}\sigma^{4}+36\pi^{2}P^{4}\sigma^{2}-4\pi^{2}P^{2}\sigma^{2}-12P^{4}+\pi^{6}\sigma^{6}\right)}{\sigma^{2}\left(6P^{2}+\pi^{2}\sigma^{2}\right)^{2}}\\
 & +\frac{5\left(6P^{2}+\pi^{2}\sigma^{2}\right)^{2}}{\sigma^{2}\left(12P^{2}\log\left(\frac{\pi\sigma}{P}\right)-12P^{2}-6P^{2}\log(4)+\pi^{2}\sigma^{2}\right)^{2}}\\
 & +\frac{2\left(18\pi^{2}P^{2}\sigma^{2}-27\pi^{2}P^{2}\sigma^{2}\log(4)+9\pi^{2}P^{2}\sigma^{2}\log(64)+12P^{2}+3\pi^{4}\sigma^{4}-\pi^{2}\sigma^{2}\right)}{\sigma^{2}\left(12P^{2}\log\left(\frac{\pi\sigma}{P}\right)-12P^{2}-6P^{2}\log(4)+\pi^{2}\sigma^{2}\right)}
\end{align*}

However, just as the $\sigma=0$ case, we have a logarithm and
therefore we can not obtain a polynomial expansion for the U function.
The conclusion is the same: we can not apply the Kovacic's criteria
for the $+_{P,N}$ theory.

\section{General Observations about Expansions and Integrability.}
In this section we study some aspects of Polylogarithmic expansions in order to understand how to apply (or not) the Kovacic's criteria.

\subsection{The truncation at $\eta=0$.}
For all the cases considered here, we have been able to study integrability with the truncation $\eta=0$. Therefore, we can try to study the general case. In order to do this we must remember that the general potential for the quivers is given by \cite{Legramandi:2021uds}
\begin{equation}\label{generalpoly}
\hat{V}=\frac{P^{2}}{2\pi^{3}}\sum_{s=1}^{P-1}c_{s}\mbox{Re}\left[\text{Li}_{3}\left(e^{-\frac{\pi(-is+i\eta+\sigma)}{P}}\right)-\text{Li}_{3}\left(e^{-\frac{\pi(is+i\eta+\sigma)}{P}}\right)\right],
\end{equation}
where
$$
c_{s}=(2N_{s}-N_{s-1}-N_{s+1}).
$$

In this case we expand in $\eta$ to get
\begin{align}
\hat{V} & =\frac{P^{2}}{2\pi^{3}}\sum_{s=1}^{P-1}c_{s}\mbox{Re}\left(\text{Li}_{3}\left(e^{-\frac{\pi(\sigma-is)}{P}}\right)-\text{Li}_{3}\left(e^{-\frac{\pi(is+\sigma)}{P}}\right)\right)\nonumber
\\
 & -\frac{P}{2\pi^{2}}\eta\sum_{s=1}^{P-1}c_{s}\mbox{Re}\left(i\text{Li}_{2}\left(e^{-\frac{\pi(\sigma-is)}{P}}\right)-i\text{Li}_{2}\left(e^{-\frac{\pi(is+\sigma)}{P}}\right)\right) \nonumber
 \\
 & +\frac{1}{4\pi}\eta^{2}\sum_{s=1}^{P-1}c_{s}\mbox{Re}\left(\log\left(1-e^{-\frac{\pi(\sigma-is)}{P}}\right)-\log\left(1-e^{-\frac{\pi(\sigma+is)}{P}}\right)\right)\nonumber
 \\
 & -\frac{1}{12P}\eta^{3}\sum_{s=1}^{P-1}c_{s}\mbox{Re}\frac{i\left(e^{\frac{\pi(-\sigma-is)}{P}}-e^{\frac{\pi(-\sigma+is)}{P}}\right)}{\left(-1+e^{\frac{\pi(-\sigma-is)}{P}}\right)\left(-1+e^{\frac{\pi(-\sigma+is)}{P}}\right)}+O\left(\eta^{4}\right).\label{generaletaexpansion}
\end{align}
With this result we arrive at
\begin{align}
\partial_{\eta}f_{3} & =\frac{\pi^{2}}{3P^{2}\sigma}\mbox{Re}\sum_{s=1}^{P-1}c_{s}\frac{i\left(\log\left(1-e^{-\frac{i\pi(s-i\sigma)}{P}}\right)-\log\left(1-e^{\frac{i\pi(s+i\sigma)}{P}}\right)\right)^{2}}{\left(\text{Li}_{2}\left(e^{\frac{i\pi(s+i\sigma)}{P}}\right)-\text{Li}_{2}\left(e^{-\frac{i\pi(s-i\sigma)}{P}}\right)\right){}^{2}}
\label{etatruncation}
\\
 &-\frac{\pi^{2}}{3P^{2}\sigma}\mbox{Re}\sum_{s=1}^{P-1}c_{s}\frac{1}{\left(\text{Li}_{2}\left(e^{\frac{i\pi(s+i\sigma)}{P}}\right)-\text{Li}_{2}\left(e^{-\frac{i\pi(s-i\sigma)}{P}}\right)\right)}\left(\frac{\sin\frac{\pi s}{P}}{\cos\frac{\pi s}{P}-e^{-\frac{\pi\sigma}{P}}}\right).\nonumber
\end{align}
If we use the property $\text{Li}_{s}(z^*)=\text{Li}^*_{s}(z)$ we find that the terms inside the sum are pure imaginary and, therefore, the right hand side is null. We conclude that (\ref{etatruncation}) is linear in $\eta$ and $\eta=0$ is a good truncation. 

Let us now turn our attention to the possibility of integrability. From the last term of (\ref{generaletaexpansion}) we see that we will in general have a logarithm contribution. In fact, all the terms give a logarithm contribution since we have for the Polylogarithm of positive integer
order 
\[
{\displaystyle \mbox{Li}_{s}(e^{\mu})=\frac{\mu^{n-1}}{(n-1)!}\left[H_{n-1}-\ln(-\mu)\right]+\sum_{k=0,k\neq n-1}^{\infty}\frac{\zeta(n-k)}{k!}\mu^{k},}
\]
where
\[
{\displaystyle H_{n}=\sum_{h=1}^{n}\frac{1}{h},\qquad H_{0}=0.}
\]
This express the fact that the Polylogarithm $\mbox{Li}_{s}(z)$ is
singular close to $z=1$. The emergence of this logarithm is problematic
since they are not polynomial and we can not apply Kovacic's criteria.
The only particular case in which this does not happens is when $s=P-1$.
This is due to the fact that $e^{-\frac{\pi(-is+i\eta)}{P}}=-e^{-\frac{\pi(i+i\eta)}{P}}$.
In this case, the Polylogarithm is polynomial around $z=-1$. The
only case in which we have an $s=P-1$ contribution is the $\tilde{T}_{N,P}$
theory, where we have $c_{s}=NP\delta_{s,P-1}$. Any other case will
contain $s=1$ and we necessarily have logarithm contributions an
in the $+_{P,N}$ case considered before. This explains the fact that,
for the quivers, we have got a polynomial U function only for the
$\tilde{T}_{N,P}$ case. We also point that we did not expand in $1/P$. Therefore the above result is valid for any quiver, such as the $+_{MN}$. 

\subsection{The Truncation at $\sigma=0$.}
We consider now general aspects of $\sigma=0$ truncation. We stress that the Type IIA Abelian and non-Abelian T-duals are not in the class discussed in the last subsection. However, as we saw before, the truncation $\eta=0$ is consistent for these cases and we were able to study integrability. In this case we must expand the potential (\ref{generalpoly}) in $\sigma$ to get
\begin{align*}
\hat{V} & =\frac{P^{2}}{2\pi^{3}}\sum_{s=1}^{P-1}c_{s}\mbox{Re}\left(\text{Li}_{3}\left(e^{-\frac{\pi(i\eta-is)}{P}}\right)-\text{Li}_{3}\left(e^{-\frac{\pi(is+i\eta)}{P}}\right)\right)\\
 & -\frac{P}{2\pi^{2}}\sigma\sum_{s=1}^{P-1}c_{s}\mbox{Re}\left(\text{Li}_{2}\left(e^{-\frac{\pi(i\eta-is)}{P}}\right)-\text{Li}_{2}\left(e^{-\frac{\pi(is+i\eta)}{P}}\right)\right)\\
 & -\frac{1}{4\pi}\sigma^{2}\sum_{s=1}^{P-1}c_{s}\mbox{Re}\left(\log\left(1-e^{-\frac{\pi(i\eta-is)}{P}}\right)-\log\left(1-e^{-\frac{\pi(i\eta+is)}{P}}\right)\right)+O\left(\sigma^{3}\right).
\end{align*}
We find that
\begin{align*}
\partial_{\eta}f_{3} & =-\frac{1}{2P}\sum_{s=1}^{P-1}c_{s}\mbox{Re}\left[\frac{\left(-1+e^{\frac{2i\pi s}{P}}\right)}{\left(-1+e^{\frac{i\pi(s-\eta)}{P}}\right)\left(-1+e^{\frac{i\pi(\eta+s)}{P}}\right)}\right]\\
 & -\frac{\pi}{2P^{2}}\sum_{s=1}^{P-1}c_{s}\mbox{Re}\left[\frac{\left(-1+e^{\frac{2i\pi\eta}{P}}\right)\left(-1+e^{\frac{2i\pi s}{P}}\right)e^{\frac{i\pi(s-\eta)}{P}}}{\left(-1+e^{\frac{i\pi(s-\eta)}{P}}\right)^{2}\left(-1+e^{\frac{i\pi(\eta+s)}{P}}\right)^{2}}\right]\sigma+O\left(\sigma^{2}\right).
\end{align*}
However 
\[
\frac{\left(-1+e^{\frac{2i\pi s}{P}}\right)}{\left(-1+e^{\frac{i\pi(s-\eta)}{P}}\right)\left(-1+e^{\frac{i\pi(\eta+s)}{P}}\right)}=2i\frac{\sin\frac{2\pi s}{P}-\sin\frac{\pi(s+\eta)}{P}-\sin\frac{\pi(s-\eta)}{P}}{\left(2-\cos\frac{\pi(s-\eta)}{P}\right)\left(2-\cos\frac{\pi(s+\eta)}{P}\right)},
\]
and therefore $\partial_{\eta}f_{3}$ is linear in $\sigma$. We conclude
that, for any theory described by the potential (\ref{generalpoly}), $\sigma=0$ is
a good truncation. We also point that we did not expand in $1/P$. Then the above result is valid for any quiver, such as the $+_{MN}$. When analysing integrability, for the cases above, we arrive at the same problem as in the $\eta=0$ case. This is the reason why, also in $\sigma=0$, we were not able to apply the Kovacic's criteria. Another important point is that the Type IIA Abelian and non-Abelian T-duals are not in the above class. As we saw before, the truncation $\sigma=\sigma_0$ is not consistent for these cases and we must use the truncation $\eta=0$.

\subsection{The Truncation at Large $\sigma_0$}

In the last subsections we saw that, despite of being consistent, the truncations $\eta=0$ and $\sigma=0$ are not good to study integrability of long quivers. The reason is the appearance of logarithm dependences in the U function and this spoils the applicability of the Kovacic's criteria. However, if we consider $\sigma=\sigma_0$ with $\sigma_0$ large, we can circumvent that difficulty.

In order to analyse the long quiver with some generality we remember
that the general potential is given by \cite{Legramandi:2021uds}
\[
\hat{V}(\sigma,\eta)=\sum_{k=1}^{\infty}a_{k}\sin\left(\frac{k\pi}{P}\eta\right)e^{-\frac{k\pi}{P}|\sigma|}
\]
with
\[
a_{k}=\frac{P^{2}}{\pi^{3}k^{3}}\sum_{s=1}^{P-1}c_{s}\sin(\frac{k\pi s}{P})
\]
where $c_{s}$ does depends on $P$. In the limit of large $\sigma/P$ limit, only the $k=1$ term survive and
we have 
$$
\hat{V}(\sigma,\eta)=a_{1}\sin\left(\frac{\pi}{P}\eta\right)e^{-\frac{\pi}{P}|\sigma|}=\frac{P^{2}}{\pi^{3}}\sin\left(\frac{\pi}{P}\eta\right)e^{-\frac{\pi}{P}|\sigma|}\sum_{s=1}^{P-1}c_{s}\sin(\frac{\pi s}{P}).
$$

Now we consider the long quiver, with large $P>>1$ 
\[
\hat{V}(\sigma,\eta)=\frac{P^{2}}{\pi^{3}}\left((\frac{\pi}{P}\eta-\frac{1}{6}(\frac{\pi}{P}\eta)^{3}\right)e^{-\frac{\pi}{P}|\sigma|}f(P),
\]
where we have defined
\begin{equation}\label{f(P)}
f(P)=\frac{P^{2}}{\pi^{3}}\sin\left(\frac{\pi}{P}\eta\right)e^{-\frac{\pi}{P}|\sigma|}\sum_{s=1}^{P-1}c_{s}\sin(\frac{\pi s}{P}).
\end{equation}

From the expression above we find
\[
\frac{1}{(f_{1}f_{3})^{2}}\partial_{\eta}f_{3}=\frac{2\left(\pi^{2}\eta^{2}-6P^{2}\right)}{9\pi^{3}P\sigma^{2}}+O\left(\frac{1}{\sigma^{3}}\right)
\]
which is null in the large $\sigma$ limit. Therefore, we can apply the method if we consider $\sigma_{0},P$ very large.
As explained before, we use the NVE (\ref{sigmaconst}) with
\[
{\cal D}=\frac{\pi^{6}\eta^{6}+60\pi^{2}\eta^{2}P^{4}+12\pi^{4}\eta^{4}P^{2}-144P^{6}}{\eta\left(\pi^{2}\eta^{2}-6P^{2}\right)\left(\pi^{4}\eta^{4}+12P^{4}\right)}+{\cal O}\left(\frac{1}{\sigma}\right)
\]
and
\[
{\cal C}=-\frac{9\left(\pi^{3}P\right)\sigma}{2\left(\pi^{2}\eta^{2}-6P^{2}\right)}-\frac{\sqrt{12\sigma}\pi^{5/2}\left(\pi^{2}\eta^{2}+6P^{2}\right)}{\pi^{4}\eta^{4}+12P^{4}}+\frac{9\left(\pi^{4}\eta^{2}-4\pi^{2}P^{2}\right)}{4\left(\pi^{2}\eta^{2}-6P^{2}\right)}+{\cal O}\left(\frac{1}{\sigma}\right).
\]

Therefore the U function is given by
\begin{align*}
U & =\frac{18\pi^{3}P\sigma}{\pi^{2}\eta^{2}-6P^{2}}+\frac{24\pi^{2}\left(\pi^{2}\eta^{2}+6P^{2}\right)}{\pi^{4}\eta^{4}+12P^{4}}\sqrt{\frac{\sigma\pi}{3}}\\
 & \frac{\left(\pi^{6}\eta^{6}+60\pi^{2}\eta^{2}P^{4}+12\pi^{4}\eta^{4}P^{2}-144P^{6}\right)^{2}}{\eta^{2}\left(\pi^{2}\eta^{2}-6P^{2}\right)^{2}\left(\pi^{4}\eta^{4}+12P^{4}\right)^{2}}-\frac{9\left(\pi^{4}\eta^{2}-4\pi^{2}P^{2}\right)}{\pi^{2}\eta^{2}-6P^{2}}\\
 & 12\frac{\pi^{6}\eta^{5}+8\pi^{4}\eta^{3}P^{2}+20\pi^{2}\eta P^{4}}{\eta\left(\pi^{2}\eta^{2}-6P^{2}\right)\left(\pi^{4}\eta^{4}+12P^{4}\right)}-\frac{8\pi^{4}\eta^{2}\left(\pi^{6}\eta^{6}+60\pi^{2}\eta^{2}P^{4}+12\pi^{4}\eta^{4}P^{2}-144P^{6}\right)}{\left(\pi^{2}\eta^{2}-6P^{2}\right)\left(\pi^{4}\eta^{4}+12P^{4}\right)^{2}}\\
 & \frac{4\pi^{2}\left(\pi^{6}\eta^{6}+60\pi^{2}\eta^{2}P^{4}+12\pi^{4}\eta^{4}P^{2}-144P^{6}\right)}{\left(\pi^{2}\eta^{2}-6P^{2}\right)^{2}\left(\pi^{4}\eta^{4}+12P^{4}\right)}-\frac{2\pi^{6}\eta^{6}+60\pi^{2}\eta^{2}P^{4}+12\pi^{4}\eta^{4}P^{2}-144P^{6}}{\eta^{2}\left(\pi^{2}\eta^{2}-6P^{2}\right)\left(\pi^{4}\eta^{4}+12P^{4}\right)}.
\end{align*}

By analysing this last $U$, we see that it does not satisfy all the three necessary Kovacic's conditions. The solution to the equation should then be non-Liouvillian. The limit above includes the $(p,q)$-$5$-Brane background studied in Ref. \cite{DHoker:2016ujz} and also any long quiver. It is important to note that $f(P)$, given in Eq. (\ref{f(P)}), determines what is the specific long quiver. However, $f(P)$ does not appear in all the expressions and particularly in the U function. Therefore, we conclude that strings propagating in any long quiver, and $(p,q)$-$5$-Branes are not integrable.

\section{Numerical Analysis.}

In this section we change the focus and turn to a numerical analysis of the string dynamical system in the $AdS_{6}$ background. The main goal here is to find signals of chaotic motion due to the non-linearity of the eom for the quivers given. In finding chaotic behaviour, we give another piece of evidence for non-integrability of the models studied in this paper. Basically, chaos can be understood by the high sensitivity of the dynamical system to the initial conditions: given two trajectories with different (but close) initial conditions, they will drastically differ from each other after time evolves. The way the system behaves shows some characteristics that can be measured. For example, the distance between trajectories with time can be computed  by the Lyapunov exponents, as we will see. Another interesting characteristic is related to the periodicity of the system. In chaotic motions periodicity is lost, and this can be seem by computing the power spectra associated to the trajectories. Despite this, by studying the Poincar\'{e} sections we can understand the behaviour of trajectories in the phase space in order to see how periodicity is lost if we, for example, increase the energy of the string soliton. All of these can be better understood in the nice reference \cite{Nunez:2018ags} (see also \cite{Ott2002}). In the discussion below we explain in more details these aspects. We show, as examples, results related to the backgrounds associated to the $+_{PN}$ and $\tilde{T}_{N,P}$ quivers described respectively by the potentials (\ref{V+PN}) and (\ref{VTNP}).

First of all, it is important to remember that the string soliton is placed at the center of the $AdS_{6}$ part of the background. In this way, its motion is restricted to the $S^{2}\times\Sigma$ part. In other words, we should study the behaviour with time of the coordinates $\chi(t)$, $\sigma(t)$ and $\eta(t)$. Furthermore, due to the need of large $P$ for a trustable quiver field theory, we let the $\eta$ coordinate runs to a large interval but still being bounded. In another words, the range of $\eta$ and $\sigma$ are the intervals $[0,P]$ and $-\infty<\sigma<\infty$ respectively. This is closely related to the discussion in the work done in \cite{Nunez:2018ags} where the coordinates evolved are bounded. Given these, we turn now attention to the study of trajectories and some of its characteristics. With this in mind we made a numerical evolution of the hamiltonian equations of motion coming from eq. (\ref{Lagrangian}).
\subsection{String Trajectories and Power Spectra.}

\begin{figure}
\centering
\begin{subfigure}{.5\textwidth}
  \centering
  \includegraphics[width=1.\linewidth]{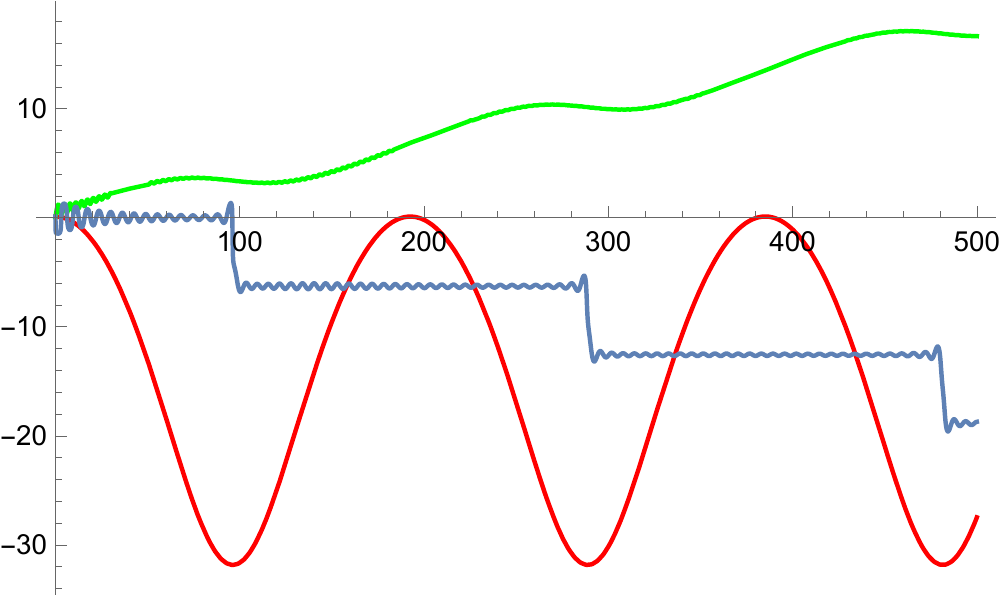}
  \caption{$E\simeq 11$.}
  \label{trajtnp1}
\end{subfigure}
\begin{subfigure}{.5\textwidth}
  \centering
  \includegraphics[width=1.\linewidth]{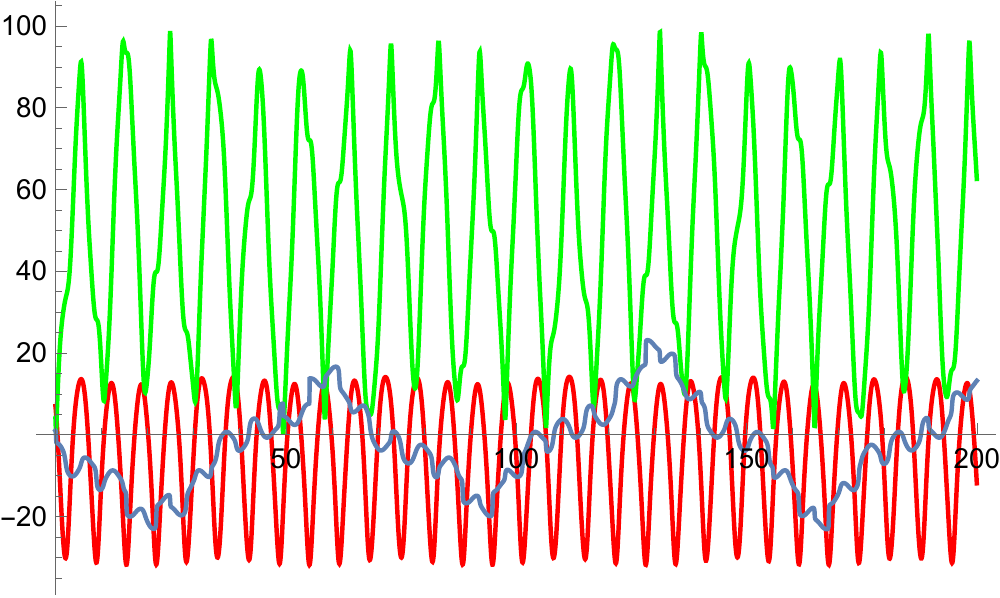}
  \caption{$E\simeq 330$.}
  \label{trajtnp2}
\end{subfigure}
\caption{String trajectories evolving in time in the case of the $\tilde{T}_{N,P}$ quiver. In red, green and blue we show respectively the $\sigma$, $\eta$ and $\chi$ coordinates. It is clear how complex they become as we increase the energy from $E\simeq 11$ to $E\simeq 330$.  }
\label{trajtnp}
\end{figure}

\begin{figure}
\begin{subfigure}{.5\textwidth}
  \centering
  \includegraphics[width=2.\linewidth]{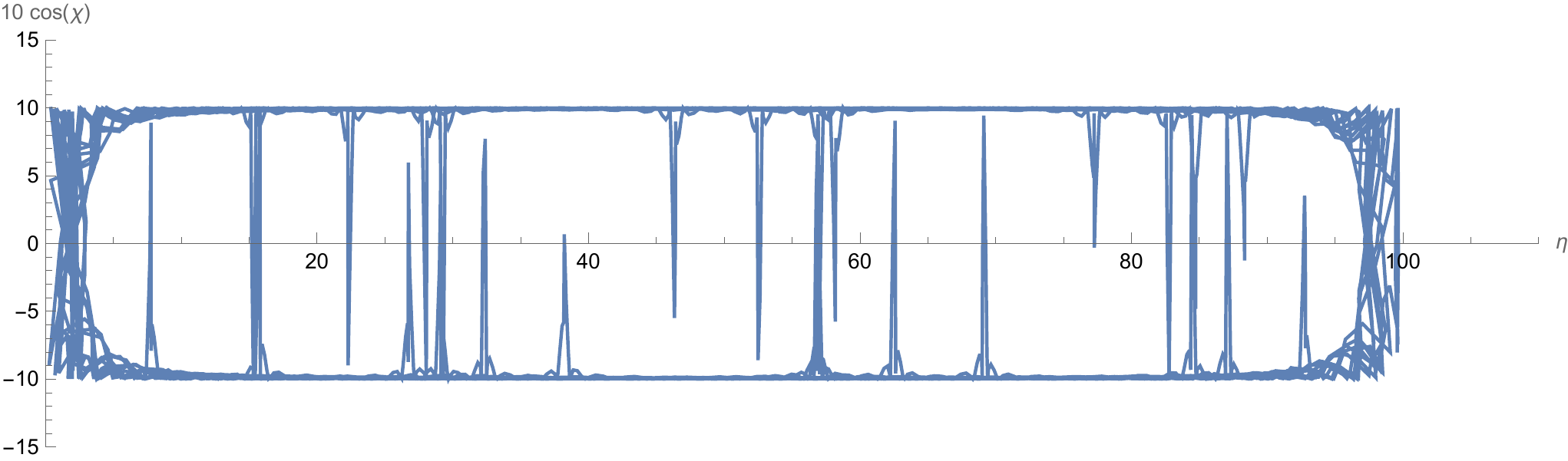}
\end{subfigure}%
\caption{In this plot we show the behavior of the compact coordinates $\chi$ and $\eta$ (with $\cos{\chi}$ rescaled by 10) with $E\simeq 11$ for the $\tilde{T}_{N,P}$ quiver.}
\label{coschieta1}
\end{figure}

\begin{figure}
\begin{subfigure}{.5\textwidth}
  \centering
  \includegraphics[width=2.\linewidth]{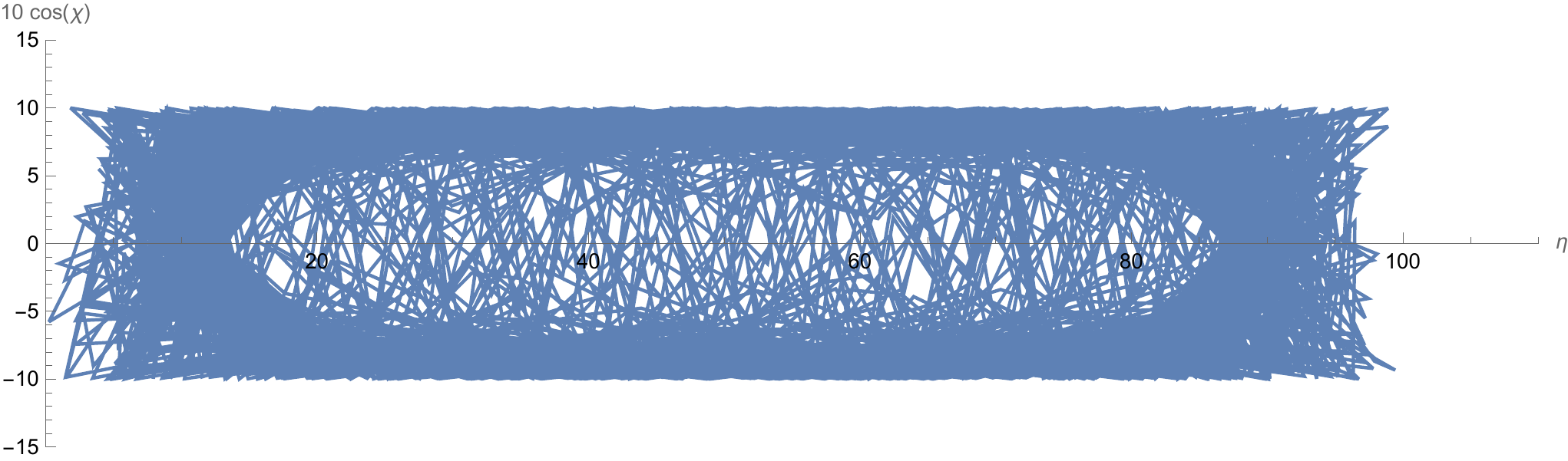}
  
\end{subfigure}
\caption{In this plot we show the behaviour of the compact coordinates $\chi$ and $\eta$ (with $\cos{\chi}$ rescaled by 10) with $E\simeq 330$ for the $\tilde{T}_{N,P}$ quiver in order to see more complex trajectories.}\label{coschieta2}
\end{figure}

Related to the trajectories, the first characteristic we call attention is to their almost periodic oscillations for low energies. In figure (\ref{trajtnp1}) we show a plot of trajectories with time (the horizontal line) for the $\tilde{T}_{N,P}$ quiver with $E\simeq 11$. We represent in red, green and blue respectively the $\sigma$, $\eta$ and $\chi$ coordinates. Despite some small oscillations, the general behaviour is almost periodic (the real non-periodicity will be shown in the subsequent discussion). If we increase the energy, then we see the trajectories in figure (\ref{trajtnp2}) with more oscillations: despite the fact that the time axes don't show the same regions, we clearly see an increasing in frequencies, for example, if we compare the plots between $t=0$ and $t=200$. Another interesting point is the change of the form of oscillations with different energies. In a periodic system we would wait change of frequencies and amplitudes, but no change of, for example, sinusoidal description of them. In order to make the numerics we chose $P=100$ in this work. The initial conditions taken for the energies are, for $E=11.04$, $\chi(0)=0.1$, $p_{\chi}(0)=0.006$, $\sigma(0)=0.1$, $p_{\sigma}(0)=0.0001$, $\eta(0)=0.03$, $p_{\eta}(0)=0.006$. For $E=330.41$, $\chi(0)=1$, $p_{\chi}(0)=2$, $\sigma(0)=7$, $p_{\sigma}(0)=2$, $\eta(0)=4$, $p_{\eta}(0)=5$. For the $+PN$ quiver we ran the initial conditions $\chi(0)=3$, $p_{\chi}(0)=5$, $\sigma(0)=3$, $p_{\sigma}(0)=5$, $\eta(0)=2$, $p_{\eta}(0)=5$ to get $E=267.18$. In this last case, we again can see basically the same trajectory behaviour for a high energetic string already present in the $\tilde{T}_{N,P}$ case.  A better way to see a chaotic signal is to plot the behaviour of the two compact coordinates $\chi$ and $\eta$ ($\chi$ is a bounded coordinate in $S^{2}$ and $\eta$ is bounded too, even for higher values, due to the quiver potential definition). We can see in figures \ref{coschieta1} and \ref{coschieta2} that characteristics for different energies for the $\tilde{T}_{N,P}$ quiver. We clearly see the string bouncing back again and again at $\eta=100$. For a high value of the energy, the trajectories in fact become more and more complex.
\begin{figure}
\centering
\begin{subfigure}{.5\textwidth}
\centering
  \includegraphics[width=1.\linewidth]{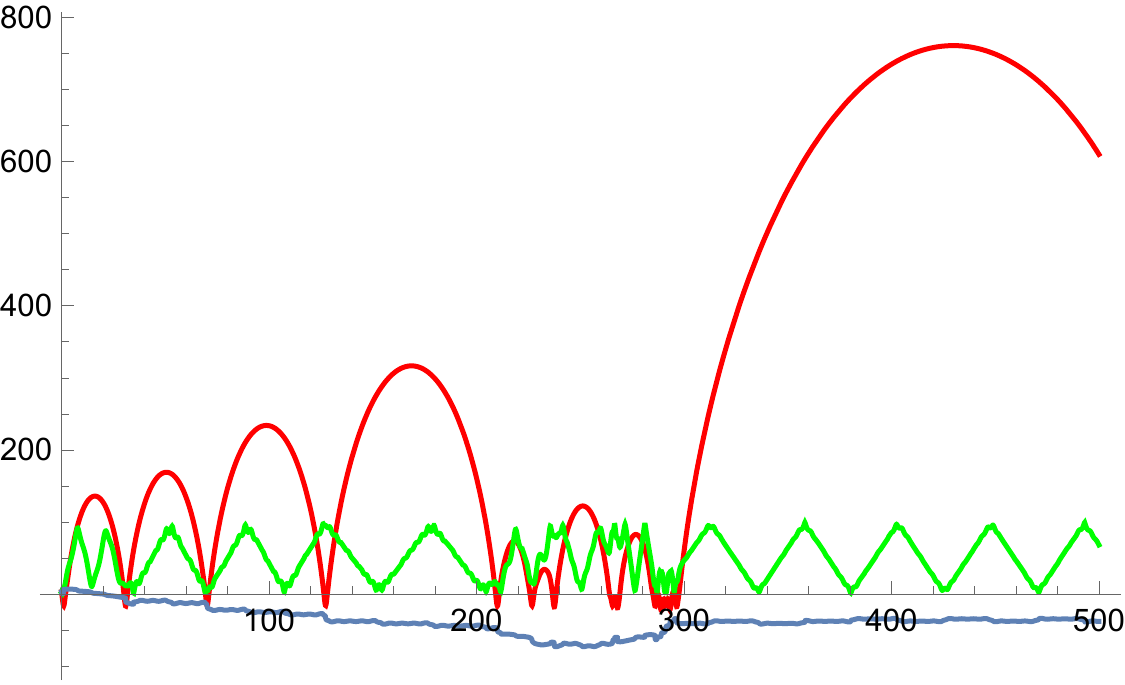}
  \caption{$E\simeq 267$.}
  \label{trajpn1}
\end{subfigure}%

\begin{subfigure}{.5\textwidth}
  \centering
  \includegraphics[width=1.\linewidth]{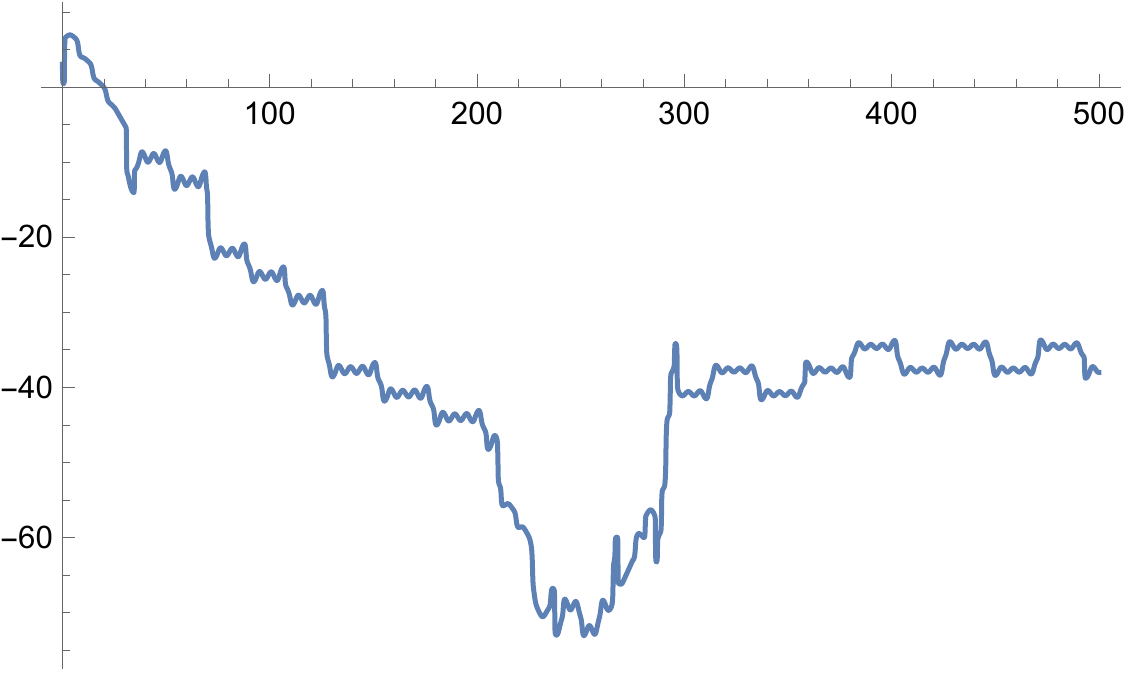}
  \caption{Plot of $\chi$ coordinate for $E\simeq 267$.}
  \label{pnch1}
\end{subfigure}
\caption{Plots with an example of trajectories for the $+_{PN}$ quiver. In this case we can see basically the same kind of behaviour present in the $\tilde{T}_{N,P}$ quiver.}
\label{trajpn}
\end{figure}


\begin{figure}
\centering
\begin{subfigure}{.5\textwidth}
  \centering
  \includegraphics[width=.96\linewidth]{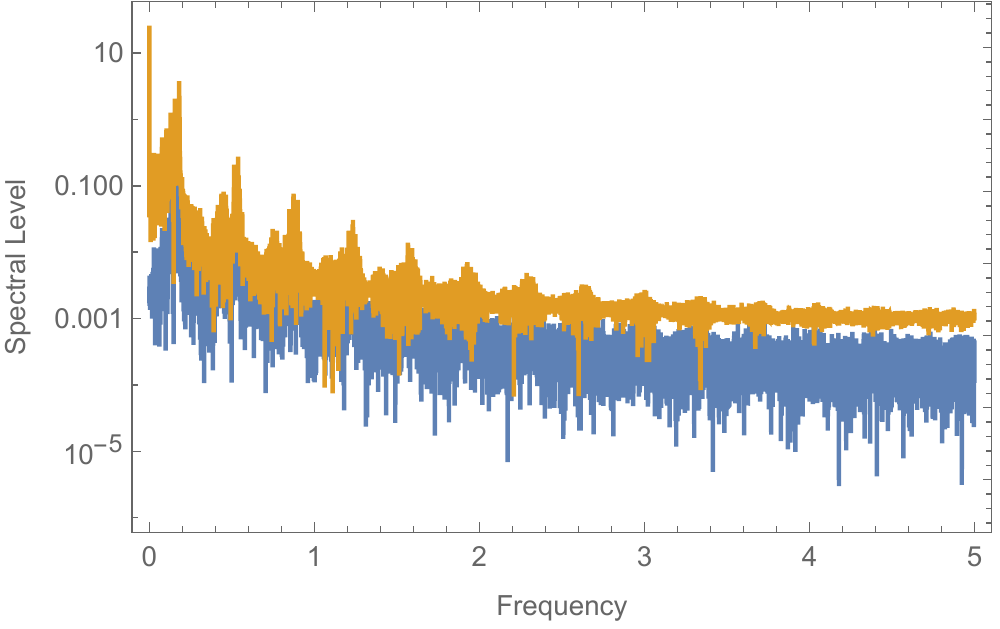}
  \caption{$E\simeq 330$ for $\tilde{T}_{N,P}$ quiver.}
  \label{powertnp}
\end{subfigure}%
\begin{subfigure}{.5\textwidth}
  \centering
  \includegraphics[width=.96\linewidth]{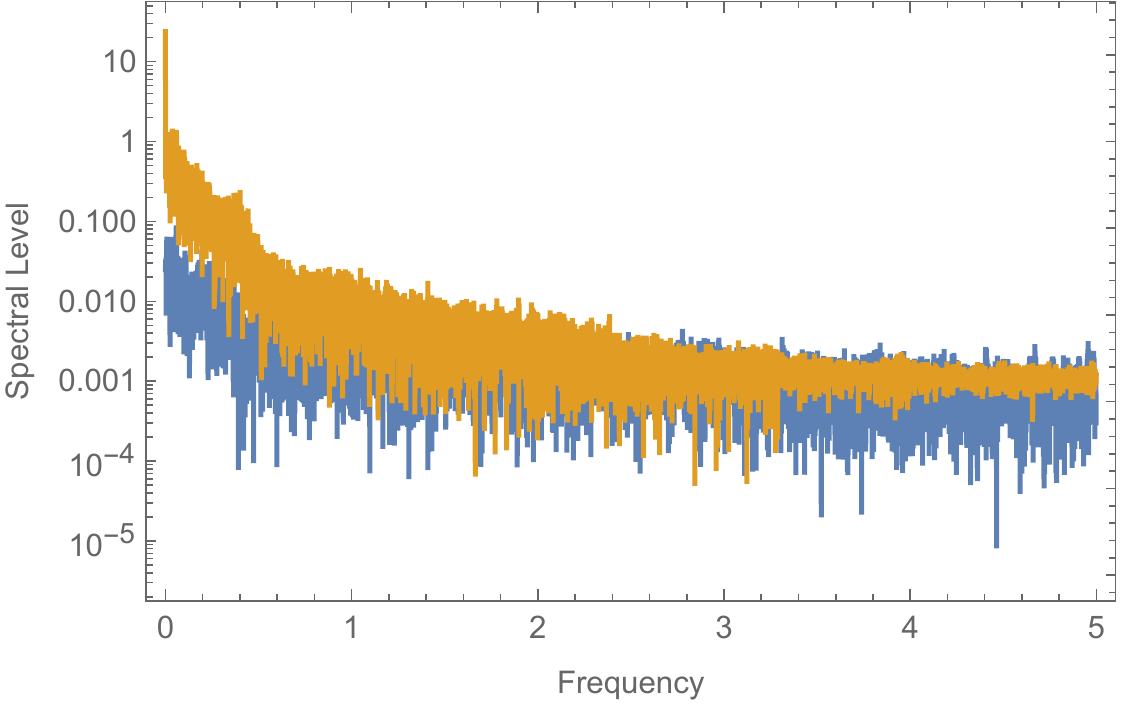}
  \caption{$E\simeq 534$ for $+PN$ quiver.}
  \label{powerpn}
\end{subfigure}
\caption{Plots of power spectra for high energetic strings in the quivers $+_{PN}$ and $\tilde{T}_{N,P}$. In blue we show the spectra for the $\chi$ coordinate and, in orange, for $\eta$. There is no principal frequency appearing in any case. The spectra are dominated by noise.}
\label{power}
\end{figure}

In order to catch non-periodicity characteristics for the string trajectories presented above, we can compute their power spectra. By taking Fourier transforms of the coordinates, the periodic, quasi-periodic or chaotic behaviour will appear. When a signal is perfectly periodic with a frequency $\omega$ its Fourier spectrum will show a vertical line at the characteristic frequency of the system. If the signal is chaotic, what we expect for high energies, the power spectra presents a noisy band of frequencies. In the figures \ref{powertnp} and \ref{powerpn} we see exactly that behaviour for the quivers $+_{PN}$ and $\tilde{T}_{N,P}$. For the $+_{PN}$ case we set the string energy to $E\simeq 534$ in which the initial conditions are given by $\chi(0)=3$, $p_{\chi}(0)=10$, $\sigma(0)=3$, $p_{\sigma}(0)=10$, $\eta(0)=2$ and $p_{\eta}(0)=10$. In this way we see clearly its noisy power spectra due to chaotic behaviour.
\subsection{Lyapunov Exponents.}

\begin{figure}[htbp]
    \includegraphics[width=.50\textwidth]{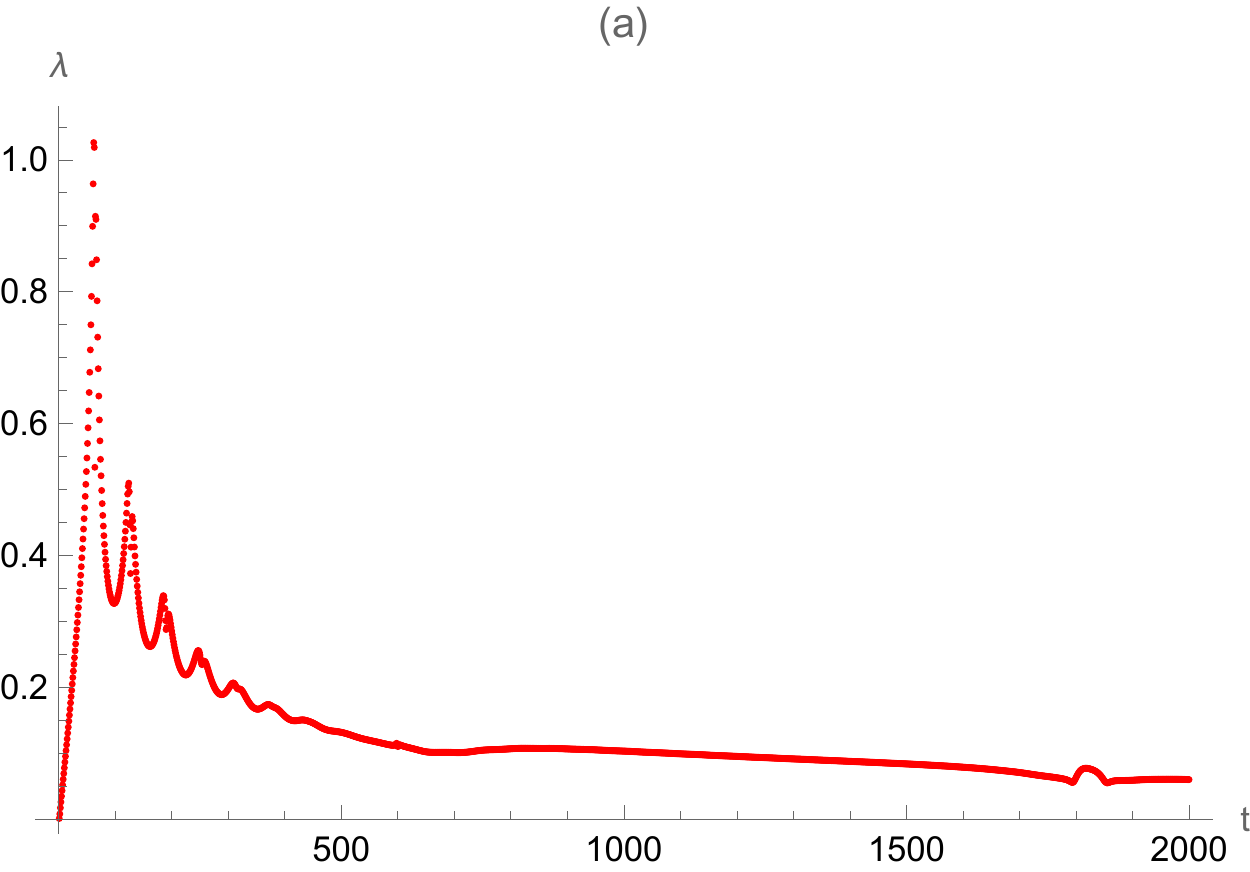}\hfill
    \includegraphics[width=.50\textwidth]{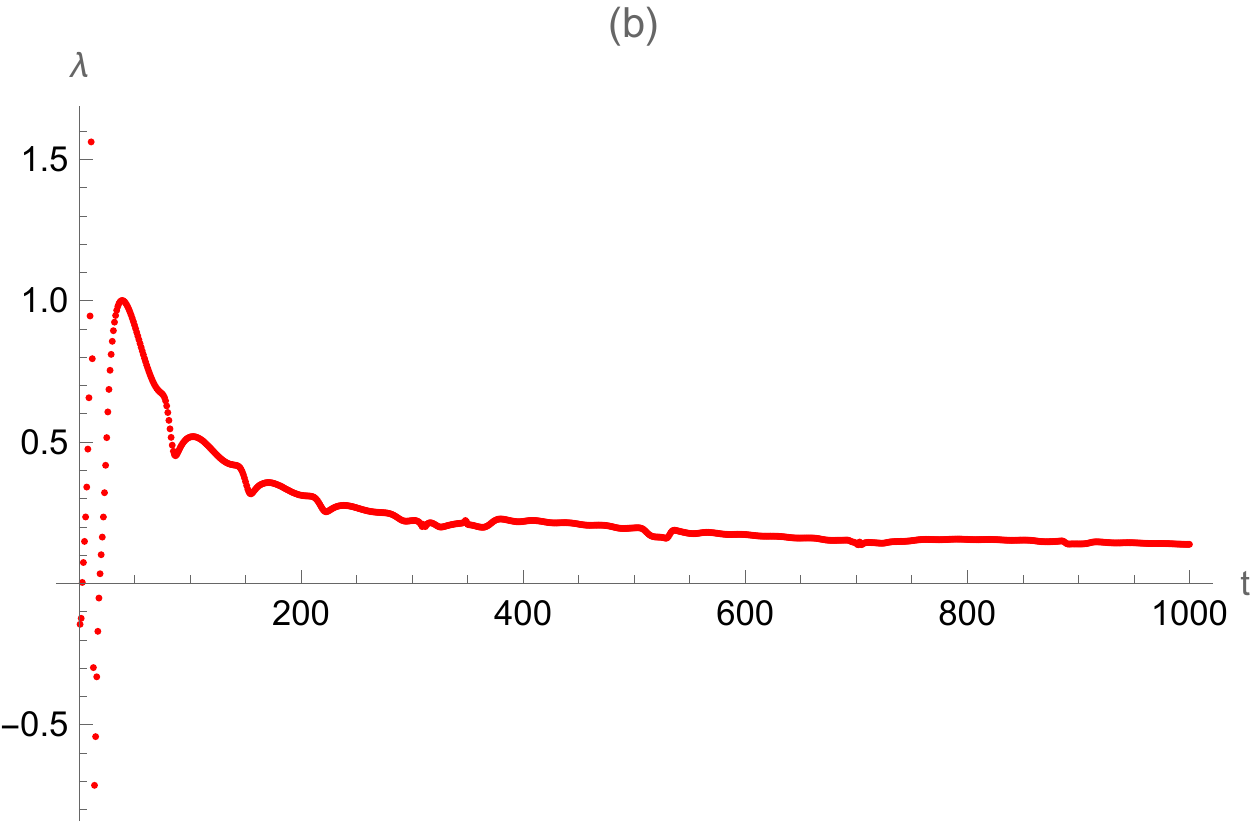}
     \caption{Plots of Lyapunov coefficients for the $+_{PN}$ quiver with different energies. We can see that for higher values of $t$ the exponential decays to a constant non-zero number. In figure (a) we have $E\simeq 15$ and for (b) $E\simeq 267$.}
     \label{lcpn}
\end{figure}

\begin{figure}[htbp]
    \includegraphics[width=.50\textwidth]{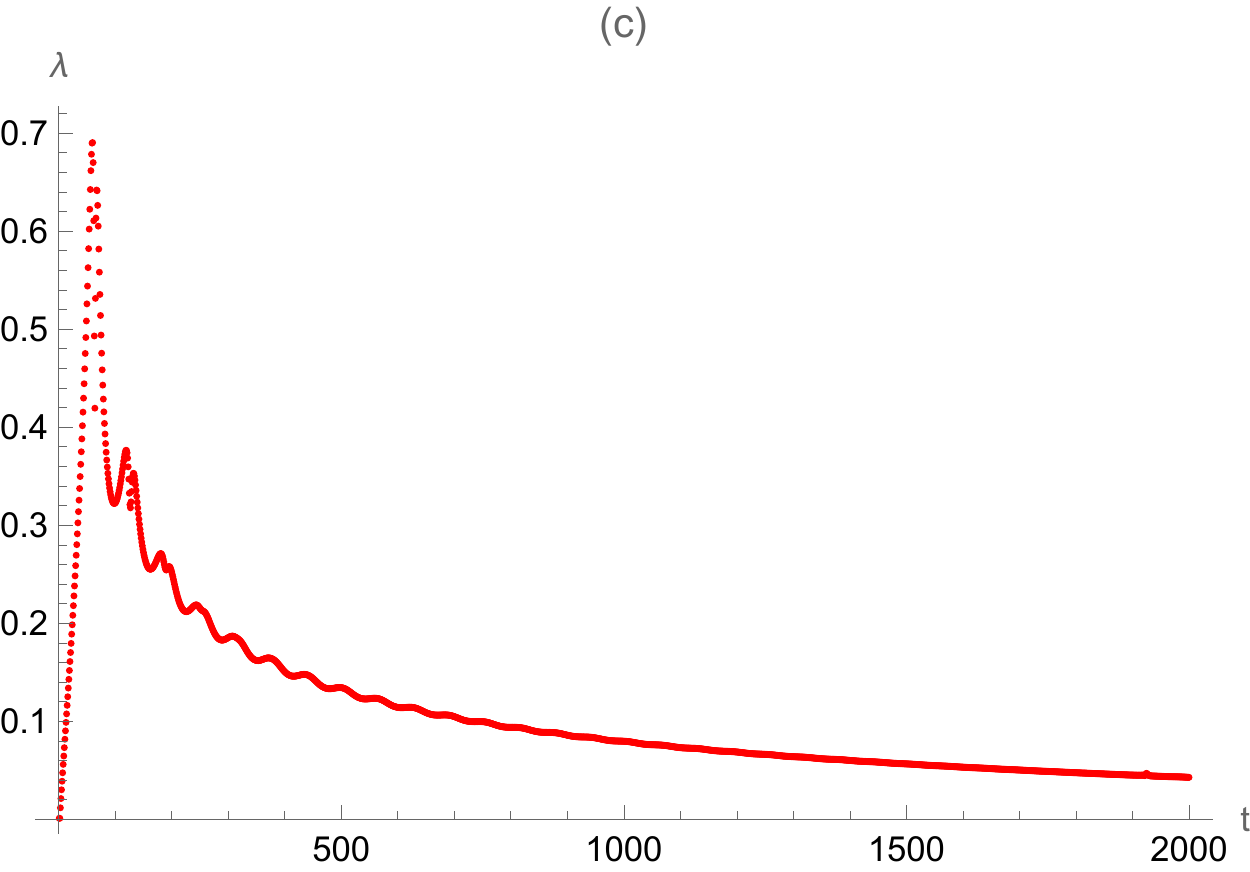}\hfill
    \includegraphics[width=.50\textwidth]{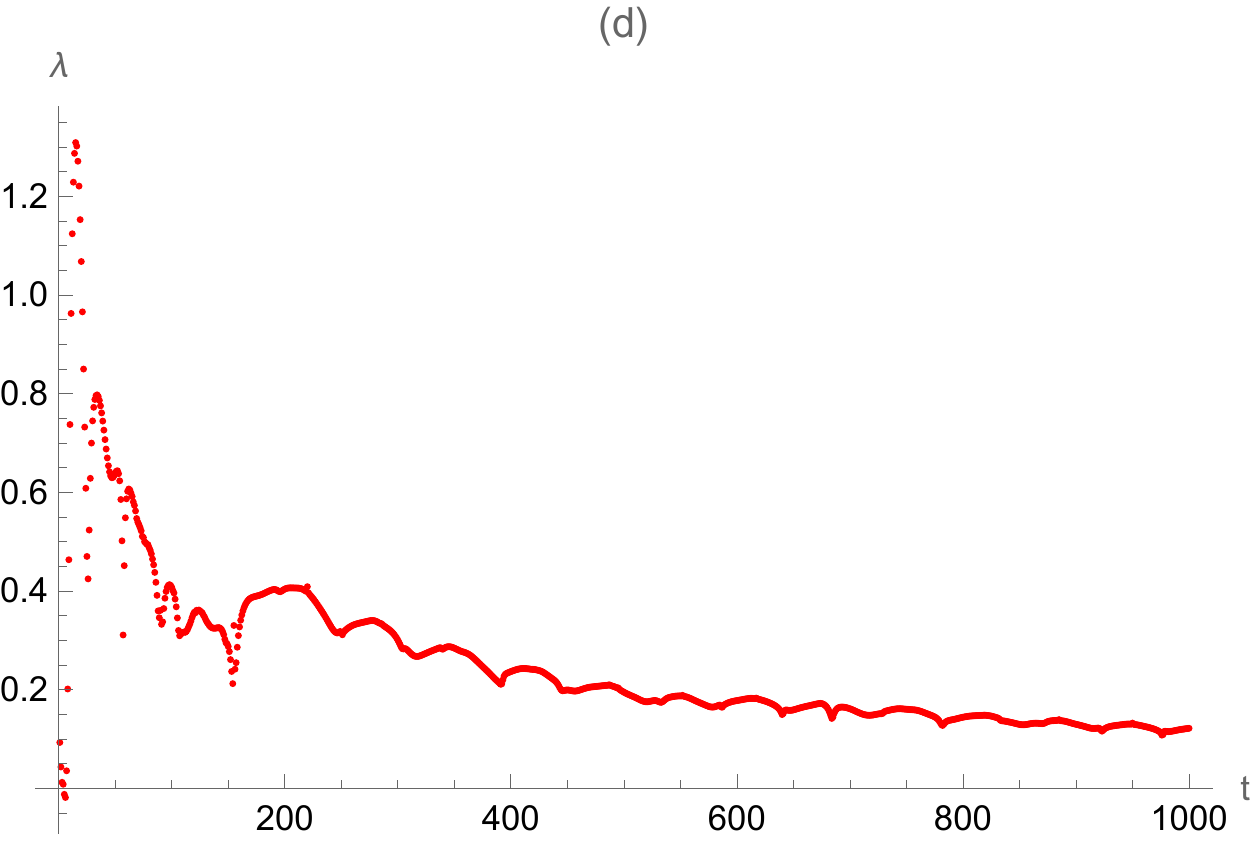}
     \caption{Plots of Lyapunov coefficients for the $\tilde{T}_{N,P}$ case with different energies.In figure (c) we have $E\simeq 11$ and for (d) $E\simeq 330$. We can see again that for higher values of $t$ the exponential decays to a constant non-zero number.}
     \label{lctnp}
\end{figure}

With the trajectories at hand, we can now search for more concrete chaos signals. We can estimate the Lyapunov exponent related to the dynamical system of the string soliton. For dynamical systems with a non-zero Lyapunov, the time evolution associated to two close trajectories in the phase space sensitively depends on the initial conditions, as we already cited. If we make a small change in initial conditions, this change will grow exponentially at sufficiently large times. In other words, the Lyapunov exponent $\lambda$, for a point $X=(q,p)$ in phase space with initial condition $X_{0}=(q(t)=0,p(t)=0)$ is given by
\begin{equation}
\lambda=\lim_{\tau\rightarrow\infty} \lim_{\Delta X_{0}\rightarrow 0}\frac{1}{\tau} log\frac{\Delta X(X_{0},\tau)}{\Delta X(X_{0},0)}.
\end{equation}
For non-zero exponent, chaotic systems typically evolves to
\begin{equation}
\Delta X(X_{0},\tau)\sim \Delta X(X_{0},0) \exp {\lambda\tau}.
\end{equation}
This means that, if the Lyapunov coefficient is non-zero, the exponential will decay to a non-zero value. We present estimations for the Lyapunov exponents obtained in figures \ref{lcpn} and \ref{lctnp}. In these plots we see the existence of non-zero exponents for the quivers $+_{PN}$ and $\tilde{T}_{N,P}$.

\subsection{Poincar\'{e} Sections.}
\begin{figure}
\centering
\begin{subfigure}{.5\textwidth}
  \centering
  \includegraphics[width=1.6\linewidth]{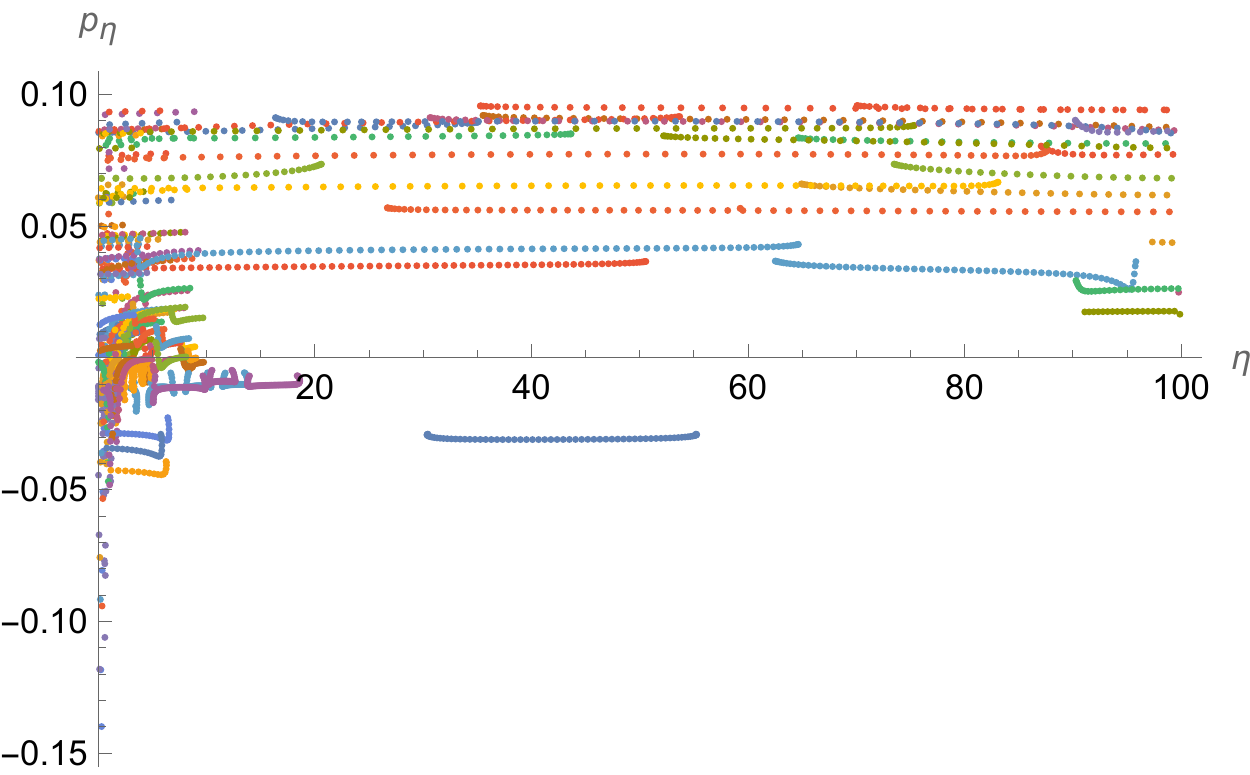}
  \caption{$E\simeq 11$.}
  \label{pstnp1}
\end{subfigure}%

\begin{subfigure}{.5\textwidth}
  \centering
  \includegraphics[width=1.5\linewidth]{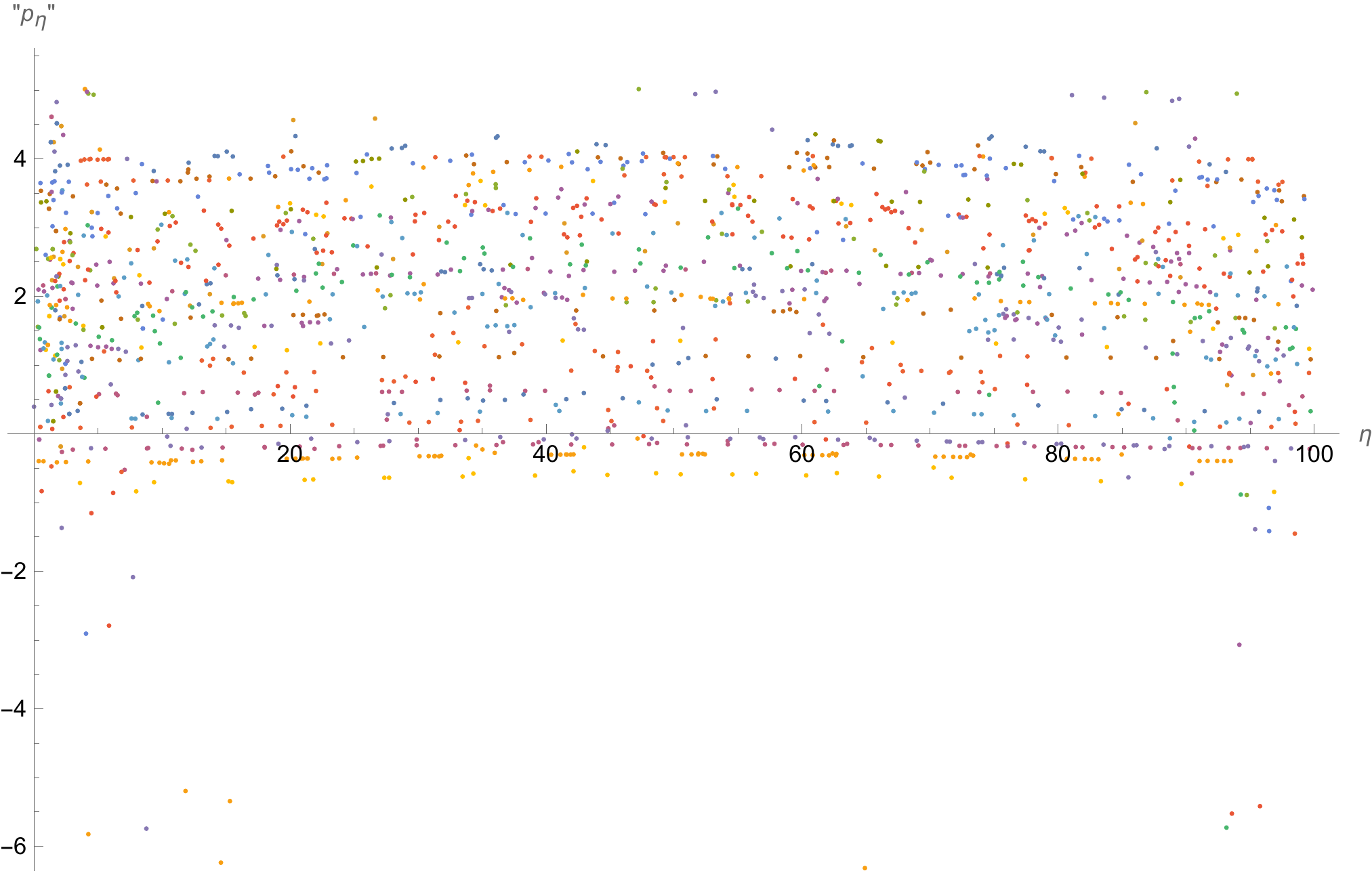}
  \caption{$E\simeq 330$.}
  \label{pstnp2}
\end{subfigure}
\caption{Poincar\'{e} sections for $\tilde{T}_{N,P}$ quiver.}
\label{pstnp}
\end{figure}

\begin{figure}
\centering
\begin{subfigure}{.5\textwidth}
  \centering
  \includegraphics[width=1.5\linewidth]{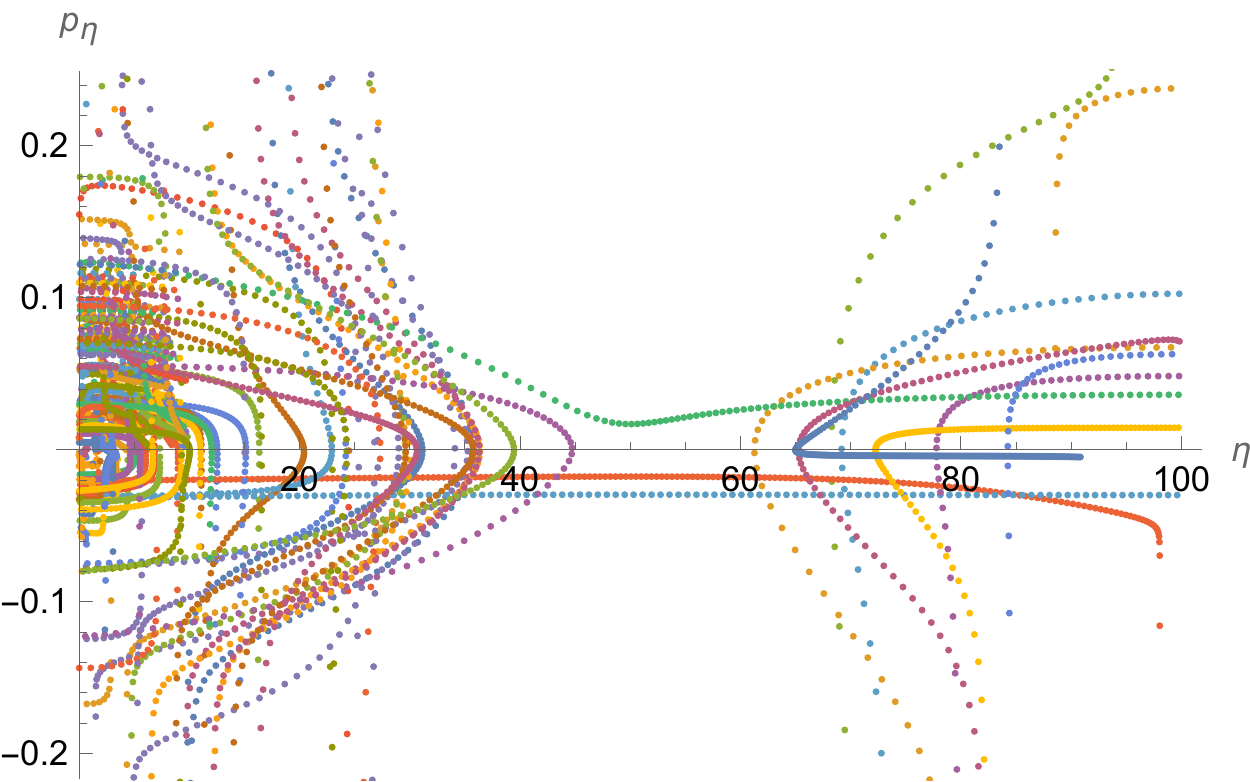}
  \caption{$E\simeq 5$.}
  \label{pspn1}
\end{subfigure}%

\begin{subfigure}{.5\textwidth}
  \centering
  \includegraphics[width=1.5\linewidth]{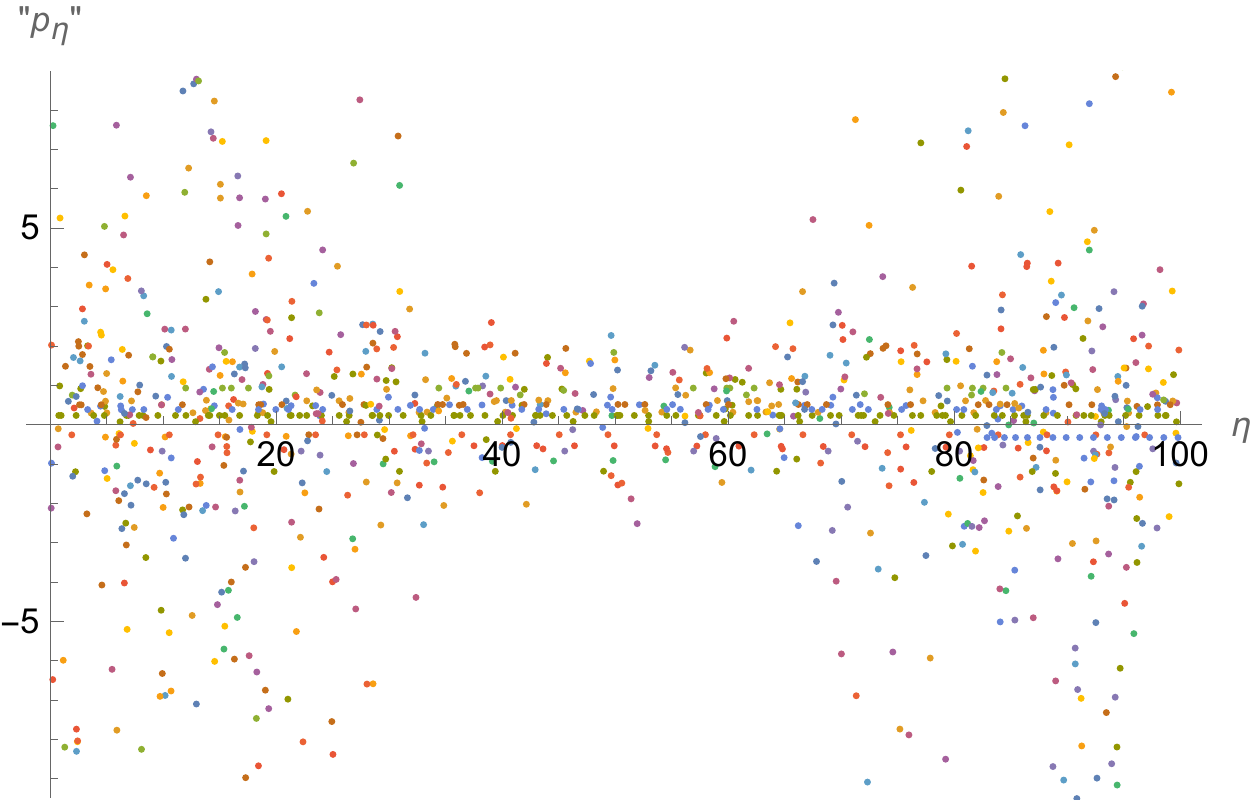}
  \caption{$E\simeq 267$.}
  \label{pspn2}
\end{subfigure}
\caption{Poincar\'{e} sections for the $+_{PN}$ quiver.}
\label{pspn}
\end{figure}

Finally we present Poincar\'{e} sections for quivers $+_{PN}$ and $\tilde{T}_{N,P}$. The analysis of the sections reveals (non)integrability of dynamical systems in the following way. An N-dimensional integrable system has N independent integrals of motion for which the Poisson bracket of any of two of them (the conserved quantities) vanishes.  Because of this, the related phase space trajectories are confined to the surface of a N-dimensional KAM torus. By taking sections of these trajectories we can see, after perturbing this torus, the change in structure of points. In a chaotic systems, the perturbation will destroy the periodic pattern in phase space. In order to generate Poincar\'{e} sections we should first generate a set of initial conditions with same energy E of the string. For example, for the case of the $+_{PN}$ quiver, for $E\simeq 267$, we made $\sigma(0)=3$ and computed random values for $p_{\sigma}(0)\in[0,10]$, $p_{\chi}(0)\in[0,10]$, $\eta(0)\in[0,20]$ and $p_{\eta}(0)\in[0,10]$. In this way we obtain several values for $\chi(0)$ such that the Virasoro constraint is obeyed for a given value of energy. For the quiver $\tilde{T}_{N,P}$, for $E\simeq 330$, we chose $\sigma(0)=7$ and computed random values for $p_{\sigma}(0)\in[0,5]$, $p_{\chi}(0)\in[0,10]$, $\eta(0)\in[0,20]$ and $p_{\eta}(0)\in[0,10]$. We plot the points $(\eta,p_{\eta})$ every time $\chi(t)=0$ in figures \ref{pstnp} and \ref{pspn}, including sections for lower energies.

In these pictures we can see how the organization of points changes if we change the string energy. For lower energies, $E\simeq 11$ for $\tilde{T}_{N,P}$ and $E\simeq 5$ for $+_{PN}$ we can see a more or less organized pattern of quasi periodic trajectories. The absence of KAM circular curves shows up for the higher energies pointed above for both quivers.  In another words, the distribution of points does not show any organization and, in fact is lost.  And these just add to the conclusion of existence of non-periodic trajectories or, in other words, to the existence of chaos.

\section{Conclusions.}

In this manuscript we studied non-integrability of strings in $AdS_{6}\times S^{2}\times\Sigma$. By using the electrostatic method \cite{Legramandi:2021uds}, we considered a very large class of theories: the long quiver, the abelian and non-abelian T-duals, the $(p,q)$-five-brane system, the $T_{N}$ and $+_{MN}$ theories. In the electrostatic method it is claimed that general dual backgrounds can be written in terms of a only function given by Eq. (\ref{solutionVhat})
\begin{equation}\label{Vgeral}
\hat{V}(\sigma,\eta)=\sum_{k=1}^{\infty}a_{k}\sin\left(\frac{k\pi}{P}\eta\right){e^{-\frac{k\pi}{P}|\sigma|}},\;\;\;\;a_{k}=\frac{1}{\pi k}\int_{0}^{P}{\cal R}(\eta)\sin\left(\frac{k\pi}{P}\eta\right)~d\eta.
\end{equation}

This method gives a clear way to link the string theory to its dual gauge theory. Of course, if the AdS/CFT conjecture is proved correct, results related to both sides of the duality will follow as consequence. With this consideration, the idea of this paper, following previous studies, is to point that the holographic duals should be (non) integrable if the string equations are (non) integrable. For this, we first carefully studied the general dynamics of strings in the background (\ref{Vgeral}) by regarding a consistent truncation of the string equations in the supergravity background. The usual procedure of the literature is to find simple solutions for $(l-1)$ of the equations, which can be replaced in the last one to obtain the NVE. However we shown that one of these equations can be replaced by the Virasoro constraint and this provides a simplification of the procedure. In this case we have three equations for $\sigma,\eta$ and $\chi$. Previous results of the literature argue that we must solve the equations for $\eta$ and $\sigma$ and replace this in the variation of $\chi$ in order to find an homogeneous second order linear equations\cite{Nunez:2018ags,Nunez:2018qcj,Filippas:2019puw}. However, depending on the background, this can be a difficult task. By using the Virasoro constraint we shown that we just have to find a simple solution for $\eta$ or $\sigma$. The fact is that  the Virasoro constraint  fixes the value of the Hamiltonian of the $1d$ system (\ref{Lagrangian}) to zero, and  the constraint $H = 0$ will always imply one of the equations of motion.
We also found a general condition to discover how simple this solution can be. We found that $\eta=\eta_0$ or $\sigma=\sigma_0$, respectively, are consistent truncations if  
\begin{equation}\label{conditionc}
\frac{1}{f_{1}^{2}f_{3}^{2}}\partial_{\eta}f_{3}|_{\eta=\eta_0}=0\mbox{ or }\frac{1}{f_{1}^{2}f_{3}^{2}}\partial_{\sigma}f_{3}|_{\sigma=\sigma_0}=0.
\end{equation}
We first applied the conditions above to look for the simple solution $\sigma=\sigma_0$. We applied it to the abelian and non-abelian T-duals, the $\tilde{T}_{N,P}$ and $+_{P,N}$ cases. We found that for the abelian and non-abelian T-duals cases $\sigma=\sigma_0$ is not a consistent truncation. For $\tilde{T}_{N,P}$ we found $\sigma=0$ a good consistent truncation, providing the applicability of Kovacic's criteria: in this case the equations of motion are non-integrable. For the $+_{P,N}$ we found again $\sigma=0$ as a good truncation, but we found logarithmic dependences on the coefficients which turn into non-applicability of the Kovacic's criteria. In summary, $\sigma=0$ is not always a good truncation, as suggested in previous studies \cite{Nunez:2018ags,Nunez:2018qcj,Filippas:2019puw}. 

Turning to the $\eta=\eta_0$ possibility of truncation we found that all cases can be studied. More specifically, $\eta=\eta_0$ is a good truncation for the abelian T-dual case, $\eta=0$ is good for the non-abelian T dual, T$_{N}$, $+_{MN}$, $\tilde{T}_{N,P}$ and $+_{P,N}$ cases. For the non-abelian T-dual case we directly applied the Kovacic's criteria finding it a non-integrable model. For the $T_{N}$ and $+_{MN}$ cases we found coefficients in non-fractional polynomial form, which do not makes possible the application of the criteria. However, we turn to the large $\sigma$ limit in which we can apply them: in this region, the criteria tell us that these models are non-integrable. For the $\tilde{T}_{N,P}$ case, as in $\sigma=0$ truncation, we found it again non-integrable, as expected. Nevertheless, in the $+_{P,N}$ we found again logarithmic dependence inside the coefficients which turn into non-applicability of the Kovacic's criteria.

Next, we considered the general case, with arbitrary potential given by (\ref{Vgeral}). We found, for $\sigma=\sigma_0$ and for $\eta=\eta_0$, a potential which is, in general, not a fractional polynomial and therefore the Kovacic's criteria cannot be applied. Nevertheless, we found the origin of logarithm contributions to the coefficients entering in the NVE equations by studying expansions of Polylogarithmic functions appearing in the general potential. In order to circumvent these contributions we shown that we can go to the large $\sigma$ and $P$ limits, finding well behaved coefficients which, consequently, gives us a nice $U-$ function to apply the criteria. In this case, by analysing the pole structure, we were able to show that it does not satisfy the Kovacic's conditions and, therefore, it is not integrable. It is interesting that in these limits we found an universal potential for long quivers. With universal we mean that $f(P)$, given in Eq. (\ref{f(P)}), determines what is the specific long quiver and it does not enter in the final expressions for the U function. This limit describes the long quivers and we conclude that all long quivers are not integrable. This includes not just the ones studied here, but all its versions, as described in in Ref. \cite{Uhlemann:2019ypp}. This includes the $+_{N,M}$, $T_N$, $Y_N$,$\ensuremath\diagup\!\!\!\!\!{+}_N$,$T_{2K,K,2}$, $T_{N,K,j}$ and $+_{N,M,j}$ theories\cite{Aharony:1997bh,Benini:2009gi,Bergman:2014kza,Hayashi:2014hfa,Bergman:2018hin,Chaney:2018gjc}.

Finally, we made a numerical treatment of the evolution of the string dynamical system. The main goal was to give more evidence of non-integrability by searching for chaotic behaviour hidden in the string eom. We plotted the trajectories for quivers $+_{PN}$ and $\tilde{T}_{N,P}$ cases, observing the increase of path's complexity for coordinates evolving in time. We pointed the sign of chaotic trajectories by the plots of the bounded coordinates $\eta$ and $\chi$, showing its behaviour for higher energetic strings. We added to this the power spectra of trajectories by showing in fact the non-existence of periodic motion for the quivers treated. After this we made computations of Lyapunov exponents for the $+_{PN}$ and $\tilde{T}_{N,P}$ cases. We concluded that the large time behaviour of trajectories contributes to non-zero values of Lyapunov exponents which signals for chaos underlined in the system. For last, we obtained its Poincar\'{e} sections. These sections points to the non-periodicity of trajectories in phase space as we increase the string energy: the more it grows, the more non-periodic is the behaviour of trajectories. This is another signal of chaos. Therefore, we add these to the non-integrability box of the $AdS_{6}$ background.

After all discussion, the main conclusion of this paper is that, for a very large class of models, we show analytically and numerically that the string equations of motion in the $AdS_{6}\times S^{2}\times\Sigma$ background are not integrable. An interesting point to be studied is the near Penrose limit behaviour of these models. In this limit the string model is solvable and quantizable. However, it is interesting to see what happens when the metric is perturbed and look for, as an example, chaotic motion and non-integrability appearing again. This question is left for a future work. 

\section*{Acknowledgements}

The authors would like to thanks Alexandra Elbakyan and sci-hub, for removing all barriers
in the way of science. We thanks Carlos Nunez, Thiago Fleury and Dmitry Melnikov for very valuable discussions and suggestions. We also thanks Marcony Silva Cunha for the help with software Mathematica. Makarius Tahim would like to dedicate this work to Maria Renata Pinheiro, Lia Maia Tahim and Ian Maia Tahim. We acknowledge
the financial support provided by the Conselho Nacional de Desenvolvimento Científico
e Tecnológico (CNPq) and Fundação Cearense de Apoio ao Desenvolvimento Científico e
Tecnológico (FUNCAP) through PRONEM PNE0112- 00085.01.00/16.

\subsubsection*{Note added.}

At the time this paper appeared as a preprint, we became aware of Mr. D. Roychowdhury's work on arXiv studying the same subject we address here \cite{Roychowdhury:2021jqt}.

\end{document}